\documentclass[11pt]{article}

\usepackage[margin=1in]{geometry}
\usepackage[colorlinks=true,urlcolor=blue,linkcolor=blue,citecolor=blue]{hyperref}
\usepackage{lipsum}
\usepackage{amsmath,amsthm,amssymb}
\usepackage{indentfirst}
\usepackage{subcaption}
\usepackage{todonotes}
\usepackage{graphicx}
\usepackage{chngcntr}
\usepackage{authblk}
\usepackage{fancyhdr}
\usepackage{natbib}
\usepackage{algorithm}          
\usepackage{algpseudocode} 
\makeatletter
\newcommand{\linelabel}[1]{%
  \begingroup
    \edef\@currentlabel{\arabic{ALC@line}}
    \label{#1}%
  \endgroup
}
\makeatother

\counterwithout{equation}{section}

\newcommand{\x}{\mathbf{x}}
\newcommand{\X}{\mathbf{X}}

\newcommand{\nul}{\emptyset}

\newcommand{\sM}{\mathcal{M}}

\title{LvD: A New Algorithm for Computing the Likelihood of a Phylogeny}

\author[1]{David Bryant\thanks{E-mail: david.bryant@otago.ac.nz}}
\author[2]{Celine Scornavacca}
\author[3]{David L. Swofford}

\affil[1]{Department of Mathematics and Statistics, University of Otago, Dunedin 9022, Aotearoa New Zealand}
\affil[2]{ISEM, CNRS, Universit\'e de Montpellier, IRD, EPHE, Montpellier, France}
\affil[3]{Florida Museum of Natural History, University of Florida, Gainesville, FL 32611, USA}

\date{}

\begin{document}
\maketitle

\begin{abstract}
There are few, if any, algorithms in statistical phylogenetics which are used more heavily than Felsenstein's 1973 pruning method for computing the likelihood of a tree. We present LvD, (Likelihood via Decomposition), an alternative to Felsenstein's algorithm based on a different decomposition of the underlying phylogeny. It works for all standard nucleotide models. The new algorithm allows updates of the likelihood calculation in worst case $O(\log n)$ time with $n$ taxa, as opposed to worst case $O(n)$ time for existing methods. In practice this leads to appreciable improvements in likelihood calculations, the extent of speed-up depending on how balanced or unbalanced the trees are. We explore implications for parallel computing, and show that the approach allows likelihoods to be computed in $O(\log n)$ parallel time per site, compared to (worst case) $O(n)$ time. We implemented and applied the algorithm to large numbers of simulated and empirical data sets and showed that these theoretical advances lead to a significant practical speed-up, although the extent of the improvement depends on how balanced the phylogenies already are.
\end{abstract}


\section{Introduction}

In statistical phylogenetics, few tools are more fundamental than Felsenstein's {\em pruning algorithm} for computing the likelihood of a phylogeny. \cite{Felsenstein04} plays down the significance of the algorithm, describing it as a simplification of earlier methods for computing likelihoods on pedigrees. Maybe so, but a simple thought experiment reveals quite clearly how important this algorithm has been and still is. The algorithm is executed for every site, every rate class, every tree, every bootstrap replicate or MCMC iteration, every Bayesian or ML analysis, every data set. If you 
suspend all the computers carrying out phylogenetic inferences at any single time, the bulk of them will probably be executing the pruning algorithm. 

Perhaps more remarkable is that, in the fifty years since it appeared, no significant modifications to the core algorithm have been published. There has been remarkable progress in the speed and capability of phylogenetic analysis software. This has, for the most part, been achieved by improved search heuristics, approximations, clever updating algorithms, and hardware. 
The gains have been achieved by reducing the number of times the pruning algorithm is called, or reducing the fraction of the phylogeny which the algorithm is applied to, rather than by redesigning the underlying algorithm. 

In this paper we describe a new algorithm for computing the likelihood of a phylogeny. The algorithm uses dynamic programming, like Felsenstein's pruning algorithm, but uses  a different structure for the computations. 
This results in an appreciable speedup with respect to the pruning algorithm, while exhibiting the same flexibility with respect to extended models and engineering speedups.

The main difference between the two algorithms can be summarized as follows.
The dynamic programming approach used in Felsenstein's algorithm decomposes the phylogeny into clades, with partial likelihoods being computed at each node in the phylogeny for the subtree rooted at that node. Our approach is to consider more general subsets of the phylogeny, which we call \emph{components}. 

We define partial likelihoods for these objects and define a {\em decomposition tree} which can be used to compute these partial likelihoods. Importantly, the decomposition tree can always be chosen to have height at most $O(\log n)$, a property which underpins all of the improvements in running time. 
Designing algorithms using this new structure leads
to an improvement from worst-case $O(n)$ time to $O(\log n)$ for updating the likelihood of a phylogeny, and a single-site, $O(\log n)$  {\em parallel time}, likelihood computation. We call this new approach {\em Likelihood via Decomposition} (LvD). 

To an extent, LvD represents a hybridisation of two main innovations. The first is the idea of defining and computing partial likelihoods for more general subsets of a phylogeny. This was first explored in the highly original paper of  \cite{SumnerCharleston10}, who employed {\em likelihood tensors} to (heuristically) maximize the gains that can be made from repeated sub-patterns in the data.  We make use of a similar idea, modified and  restricted to make the approach computationally tractable.

The second  is the structure and properties of the decomposition tree. This is based on data structures used in \cite{BryantScornavacca19} to compute the path-length metric between trees in $O(n \log n)$ time, which in turn were derived from the algorithm of \cite{Brodal04} for computing the quartet metric. 

The structure of this paper is as follows.
\begin{itemize}
\item In Section~\ref{sec:FelsensteinReview} we review the standard pruning algorithm for computing the likelihood of a phylogeny. At the same time we introduce notation used throughout the manuscript.
\item In Section~\ref{sec:pruning} we describe the new approach in detail. The key concept here is a novel decomposition of the phylogeny giving a new method for evaluating likelihoods. We obtain a new tree structure to compute the partial likelihoods on, one that has a logarithmic height. We then outline two ways in which we can take advantage of the new structures: a faster (memory efficient) algorithm for computing likelihoods and a new theoretical bound on the time taken to compute likelihoods in parallel.
\item In Section~\ref{sec:experimental} we investigate the practical speed-ups resulting from the new updating algorithm. We base our analysis on hundreds of existing simulated and real data sets.
\item Section~\ref{sec:discussion} reviews the main contributions of the paper and sketches avenues for future work.
\end{itemize}

\bigskip

\section{Revisiting Felsenstein's pruning algorithm} \label{sec:FelsensteinReview}

\subsection{Notation}

Suppose we are given an alignment $A$, a phylogeny with branch lengths, and the parameters of an evolutionary model. Our goal is to compute the probability of observing the alignment given these parameters. We assume that sites are independent, so this probability can be computed independently for each site $A_j$. The pruning algorithm of \cite{Felsenstein73} permits to compute the probability of observing a single site in the alignment, given the phylogeny, branch lengths and model parameters. We review the algorithm and introduce  notation we will need later on.

\begin{figure}[ht]
\centerline{\includegraphics[width=0.5\textwidth]{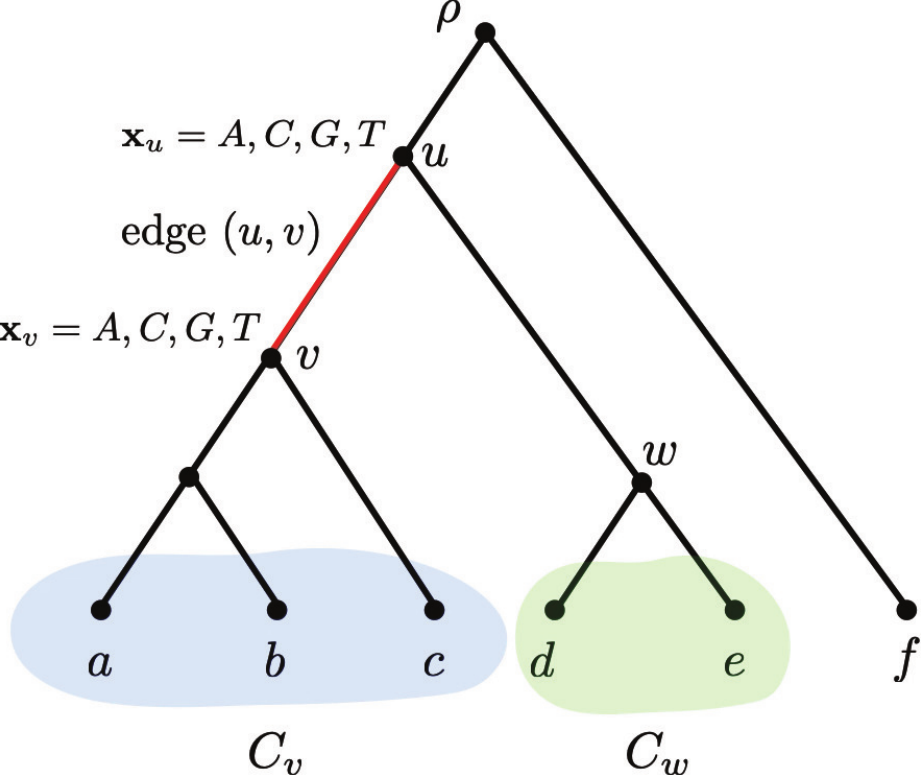}}
\caption{\footnotesize Terminology and notation for phylogenies. The root is denoted $\rho$. For each node $v$, the state at that node is denoted $\x_v$ and the set of taxa which are descendants of $v$ is denoted by $C_v$.}
\label{fig:treename}
\end{figure}

A phylogeny is made up of nodes and edges. The root is labelled  $\rho$. A node $v$ is a descendant of node $u$ if the path from $v$ to the root passes through $u$. The set of tips which are descendants of $v$ is denoted $C_v$ (Fig.~\ref{fig:treename}).

The ancestral state for node $v$ is denoted $\x_v$. The ancestral state $\x_\rho$ for the root has distribution $\pi$, defined by the evolutionary model. The joint distribution of all ancestral states is determined by the root distribution together with all the conditional probabilities 
\begin{equation}    \Pr[\x_v = s| \x_u = r]   \end{equation}   
for the state $\x_v$ of a child node $v$ given the state $\x_u$ of its parent node. These probabilities are typically functions of the edge length, according to the given model of sequence evolution. 

\subsection{Computing likelihoods using the pruning algorithm}

The pruning algorithm makes two fundamental assumptions about the model of sequence evolution:
\begin{enumerate}
\item[(i)] If we specify  (condition on) the state at a node then the  states for all descendants of one child are independent of the  states for all descendants of the other child;
\item[(ii)] If we specify (condition on) the state at a node then the  states for descendants of that node are independent of the  states for 
the ancestors.
\end{enumerate}
For any node, we let $\X_v$  denote the joint random vector containing all the states associated to nodes in the subtree $C_v$, the set of tips which are descendants of $v$.  That is, $\x_v$ is the state at the node while $\X_v$ is a random array containing the states at the tips which are descendants of $v$.

Let $u$ be a node with two children $v$ and $w$. It follows from assumption~(i) that 
\begin{equation}
\Pr[\X_v,\X_w | \x_u] = \Pr[\X_v | \x_u] \Pr[\X_w|\x_u]  \label{eq:siblings}
\end{equation}
while assumption~(ii) gives rise to identities like 
\begin{equation}
\Pr[\X_v|\x_v,\x_u] = \Pr[\X_v|\x_v]. \label{eq:vertical}
\end{equation}

The marginal probability of an alignment column $A_j$ is defined by summing over all possible assignments of states to the internal (non-tip) nodes.  The pruning algorithm of  \cite{Felsenstein73} computes this marginal probability by decomposing the probability calculation into partial likelihoods for subtrees or clades. 

We define the {\em partial likelihood} at node $v$ as the function $L_v$, with 
\begin{equation}
L_v(s) = \Pr[\X_v = \mbox{observed tip values} | \x_v = s]
\end{equation}
for all states $s$.  The interpretation is that $L_v(s)$ equals the probability that all the tips which are descendants of $v$ take on those states in the data (alignment column) conditional on node $v$ having state $s$. 

\begin{figure}[ht]
\centerline{\includegraphics[width=0.5\textwidth]{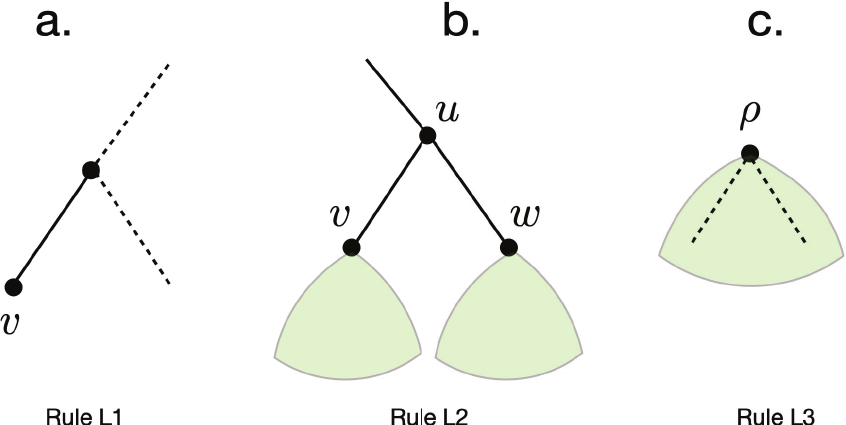}}
\caption{\footnotesize Illustration of the three rules used to evaluate the partial likelihoods for nodes in the phylogeny, using Felsenstein's algorithm: (a.) leaves; (b.) internal nodes; (c.) the root.}
\label{fig:threerules}
\end{figure}

The standard pruning algorithm is built out of three rules for computing these partial likelihoods.
\begin{enumerate}
\item[L1:] If $v$ is a tip (Fig.~\ref{fig:threerules}a) then 
\begin{equation}   L_v(s) = \begin{cases} 1 & \mbox{observed state $s$ at this tip for $A_j$;} \\ 0 & \mbox{otherwise.} \end{cases}    \end{equation}   
If the state at $v$ is missing in $A_j$, then $L_v(s) = 1$ for all $s$.
\item[L2:] If $u$ is a node with two children $v$ and $w$  (Fig.~\ref{fig:threerules}b) and  $r$ is the ancestral state at $u$ then from the conditional independence of the children we have
\begin{align}
L_u(r) & = \Pr[\X_u = \mbox{observed values } | \x_u = r] \nonumber \\ 
& = \left(\sum_s  \Pr[\x_v = s|\x_u = r] L_v(s) \right) \\ & \quad \quad \times  \left(\sum_s  \Pr[\x_w = s|\x_u = r] L_w(s) \right) .
\end{align}
\item[L3:] If $v = \rho$, the root then the probability of generating all of the states for the tips is 
\begin{equation}
\mbox{Probability of $A_j$}  = \sum_r \pi(r)  L_\rho(r).\label{eq:rootProb}
\end{equation}
where we recall that $\pi$ denotes the probabilities for states at the root. 
\end{enumerate}

The pruning algorithm starts at the tips with the first rule, works towards the root using the second rule and then evaluates the site probability using the third rule.

\section{Computing Likelihood via Decomposition Trees} \label{sec:pruning}

In this section, we present our new algorithm, define the required notation, and discuss two approaches for improving the computational efficiency of likelihood evaluation. 

\subsection{Components and their partial likelihoods}

\begin{figure}[ht]
\centerline{\includegraphics[width=0.6\textwidth]{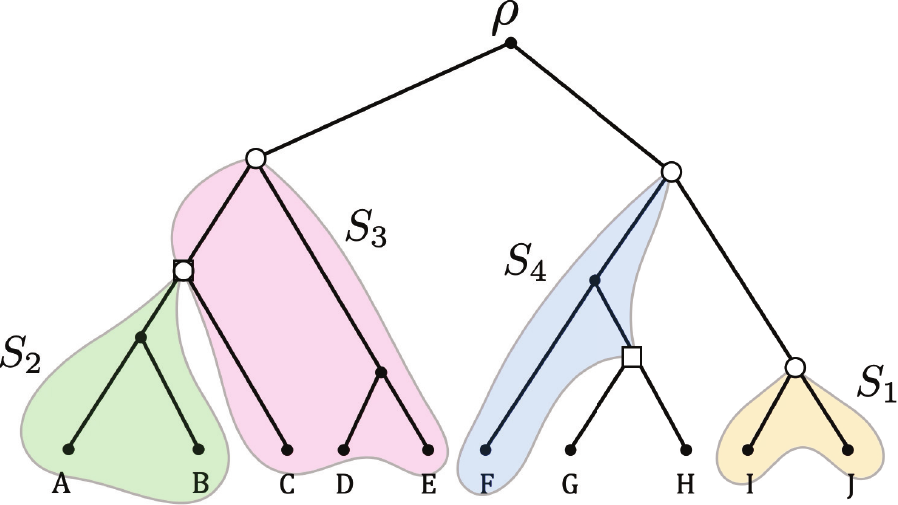}}
\caption{\footnotesize Examples of components: two clades ($S_1$ and $S_2$) and two segments ($S_3$ and $S_4$) in a tree. Boundary nodes are depicted using hollow circles and squares: hollow circles indicate component roots and hollow squares indicate component buds. Note that the root of $S_2$ coincides with the bud of $S_3$.}
\label{fig:segmentsClades}
\end{figure}

Our algorithm for computing the likelihood of a phylogeny works by extending the concept of a partial likelihood beyond subtrees to more general structures  within the phylogeny, which we now introduce. We use the term {\em component} to refer to any non-empty set of edges in a phylogeny which is {\em connected}, that is any two edges in a component are joined by a path formed by edges in that component. There is no requirement that the component contains all the descendants of nodes in that component, just that  the edges in the component are connected in the phylogeny. (Fig.~\ref{fig:segmentsClades}).

A node is a {\em boundary} of the component if it is adjacent to at least one edge inside the component and at least one edge outside the component. For convenience, we say that the root $\rho$ is always a boundary node for the component it is in. 

A component with a single boundary is called a {\em clade}. A clade contains all the edges along paths connecting a node $v$ to its descendants, possibly together with the edge connecting the node $v$ to its parent. (Fig.~\ref{fig:segmentsClades}). The single boundary node of a clade will also be the node in the clade closest to the root of the phylogeny, so we call it the {\em root} of the clade.

A component with two boundaries is called a {\em segment}. The boundary node closest to the root of the phylogeny is called the {\em root} of the segment; the other boundary node is called the {\em bud}  (Fig.~\ref{fig:segmentsClades}). 

Any segment or clade $S$ which is not just an internal edge will contain tips. We use $C_S$ to denote the (possibly empty) set of tips in a component $S$. As in the previous section, $\X_S$  shall denote the array of (random) states for the nodes in $C_S$. We can now define partial likelihoods for clades and segments. 

\begin{itemize}
\item If $S$ is a clade with root $u$ then the partial likelihood for $S$ equals function given by
\begin{equation}   L_S(r) = \Pr[\X_S = \mbox{observed states}|\x_u = r].   \end{equation}   
This is the same as the standard partial likelihood, see rule L2 above.
\item If $S$ is a segment with root $u$ and bud $v$ then the partial likelihood is the two parameter function 
\begin{equation}   \hspace{-0.5cm}L_S(r,s) = \Pr[\X_S = \mbox{observed states  and } \x_v = s|\x_u = r].  \end{equation}   
\end{itemize}
This partial likelihood is new: it includes the probability of the states at the tips and the probability of the state at the bud. It is also fundamental to our approach.

\subsection{Decomposition trees}

\begin{figure*}[htp]
\centerline{\includegraphics[width=0.9\textwidth]{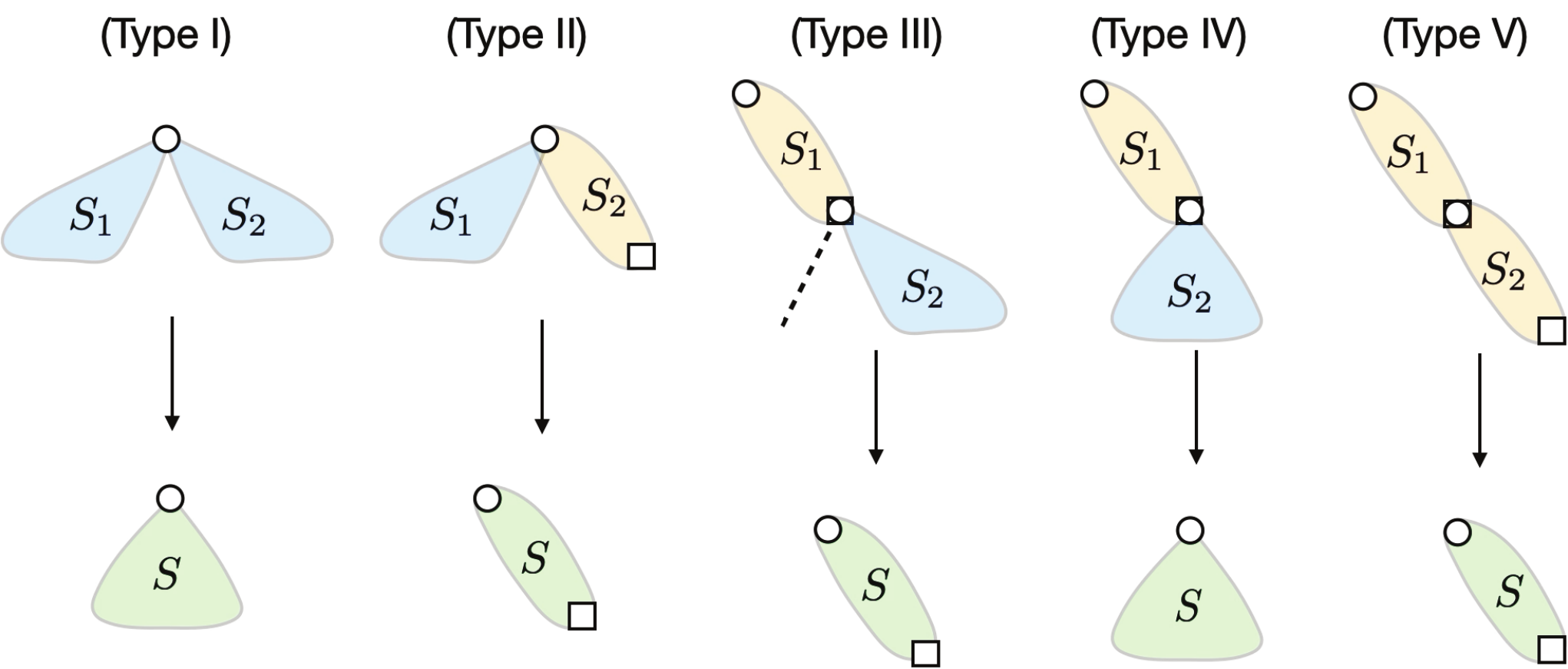}}
\caption{\footnotesize The five ways that two components ($S_1$ and $S_2$)  can be merged to give a single component ($S$).
}
\label{fig:allMergers}
\end{figure*}

The next key idea is that large components can be formed from the union of two smaller components. For this to work, the two components need to share a boundary node. As well, the resulting component has to be either a clade (with one boundary node) or a segment (with two boundary nodes). There are only a few ways that this can happen and (up to symmetry) they are all depicted in Figure~\ref{fig:allMergers}.

The smallest components are those containing a single edge. Our method works by first combining some pairs of edges into clades or segments with two edges, and continuing the process until there is one clade containing all edges. 

The sequence of mergers gives rise to a tree, which we call a {\em decomposition tree}, which satisfies the following three conditions:
\begin{enumerate}
\item Every node in the decomposition tree corresponds to a clade or segment in the phylogeny, that is, components with one or two boundaries;
\item The leaves of the decomposition tree correspond to the minimal components: the edges in the phylogeny;
\item The component corresponding to an internal node in the decomposition tree equals the disjoint union of the components for its children.
\end{enumerate}

\begin{figure*}[htb]
\centerline{\includegraphics[width=0.8\textwidth]{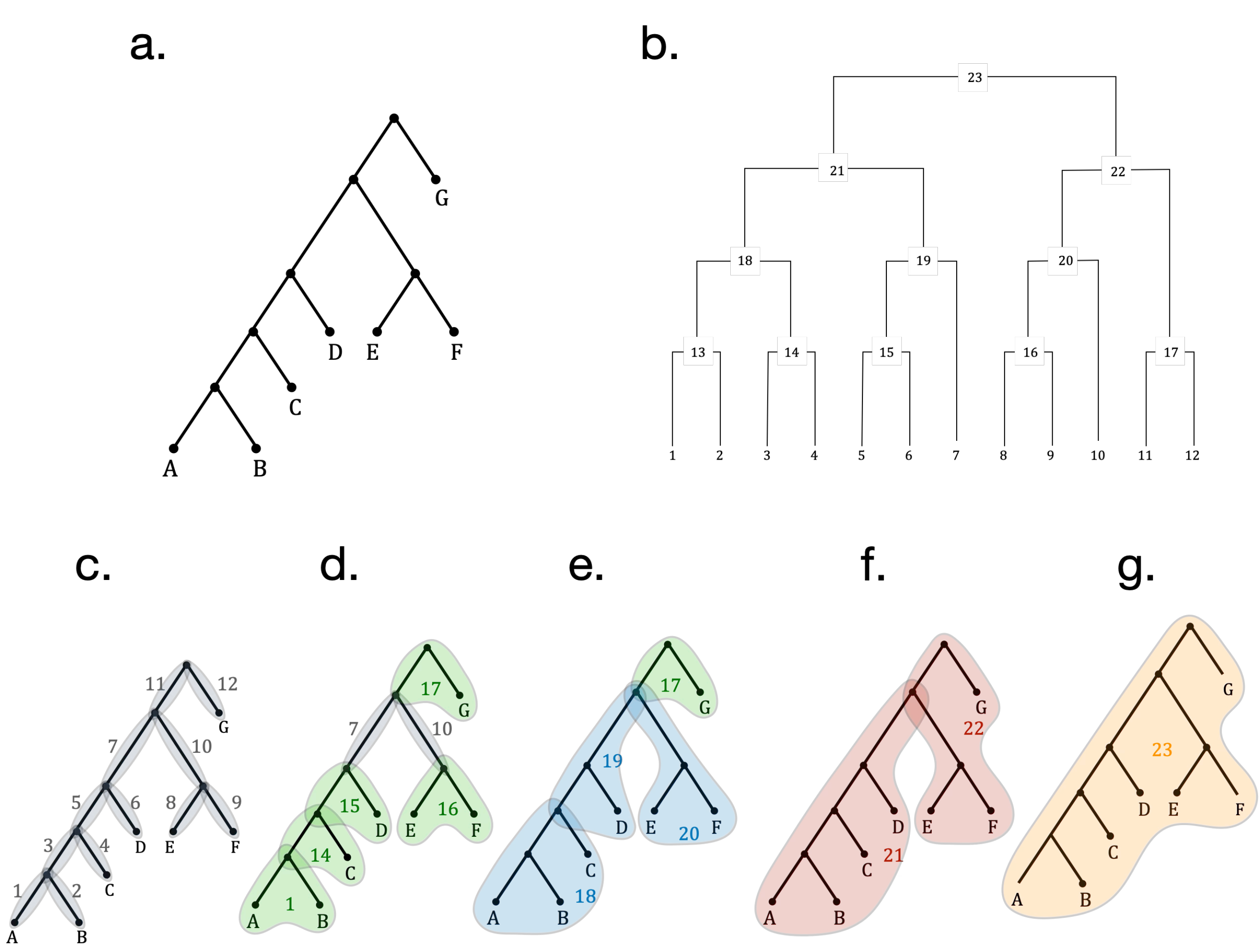}}
\caption{\footnotesize \label{fig:decomposition7taxa} A decomposition tree. (a) The original phylogeny (b) Decomposition tree (with five levels). (c) - (g) Components corresponding to each level in the decomposition tree. (c) Level one in the decomposition tree corresponds to single-edge components. Subsequent levels (d)--(g) are formed by merging components from the previous level. Note that components 7 and 10 in (c) are not merged in (d) as this would create a component with more than two boundaries.}
\end{figure*}

One example of a decomposition tree is depicted in Figure~\ref{fig:decomposition7taxa}, broken up into steps to show how large components are formed from smaller components. 

Algorithm~\ref{algo:decomp} constructs a decomposition tree. The leaves of the decomposition tree correspond to individual edges in the phylogeny. The internal nodes correspond to clades or segments formed by the union of other clades or segments using one of the five possible types of merger depicted in Fig.~\ref{fig:allMergers}. If the original (rooted) phylogeny has $n$ taxa, then the decomposition tree will have $2n-2$ leaves, one for every edge in the phylogeny. The decomposition tree is binary, so with $2n-2$ leaves it has $2(2n-2) - 1 = 4n-5$ nodes overall.

\begin{algorithm}[h!]
\small
\begin{algorithmic}[1]
\Procedure{\sc DecompositionTree}{$T$}
\State Initialize the decomposition tree $T_D$ with one node for each non-root edge in $T$
\State Initialize arrays $S,L$ of length $4n-5$
\State Fill in entries $1,\ldots,m=2n-2$ of $S$ with the single-edge components
\State $L[1:m] \gets 1$
\State $M \gets \{1,\ldots,m\}$
\State $k \gets m$
\For {$j$ from $1$ to $4n-6$} \label{line_a}
	\If{there is $i<j$ such that $i \in M$ and $S[i]\cup S[j]$ is a clade or segment} 
		\State $k \gets k+1$
		\State Create a new node $k$ in $T_D$ with children $i$ and $j$
		\State $S[k] \gets S[i] \cup S[j]$
		\State $L[k] \gets L[j]+1$
		\State $M \gets M \setminus \{i,j\} \cup \{k\}$
	\EndIf 
\EndFor \label{line-b}
\State \Return $T_D$
\EndProcedure
\end{algorithmic}
\caption{\footnotesize \label{algo:decomp} Constructing the decomposition tree for a phylogeny $T$ with $n$ leaves.}
\end{algorithm}

We construct the decomposition tree once for each phylogeny and then use the same decomposition tree for all sites of the alignment.

There are two important features of Algorithm~\ref{algo:decomp}:
\begin{enumerate}
\item The algorithm returns a valid decomposition tree, in linear time (in the number of taxa).
\item The decomposition tree constructed has  $O( \log n)$ height, irrespective of the shape of the input phylogeny. 
\end{enumerate}
We sketch proofs of these statements in the appendix. 

Note that there are many possible decomposition trees for the same phylogeny. We expect that different kinds of computation will be better served by tweaking the way that decomposition trees are constructed. Felsenstein's pruning algorithm  corresponds to a decomposition tree in which the components for internal nodes are all clades. The decomposition trees returned by Algorithm~\ref{algo:decomp} are balanced, with $O(\log n)$ height. If the input trees are already balanced then the decomposition trees will be essentially the same with either algorithm.

\subsection{Practical reduction in decomposition tree height\label{sec:expHeight}}

The theoretical bound on the height of decomposition trees constructed using Algorithm~\ref{algo:decomp} translates to a reduction in height compared to the original phylogeny, though the extent of that reduction depends on the extent to which the input phylogenies are already balanced.

\begin{figure}[htb]
\centerline{\includegraphics[width=0.8\textwidth]{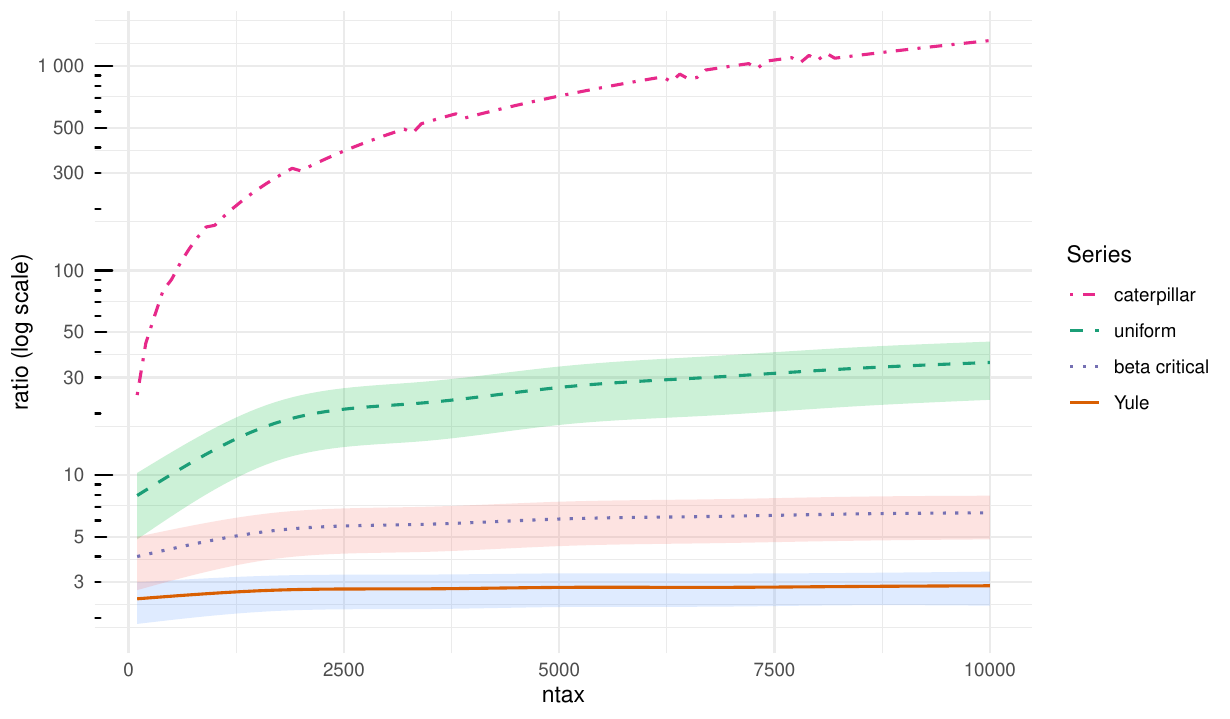}}
\caption{\footnotesize \label{fig:treeHeights} A (log-scale) plot of the height of the original tree divided by the height of the decomposition tree produced by Algorithm~\ref{algo:decomp}. 
Trees were generated from four different distributions: random caterpillar trees (completely unbalanced), random uniform trees, beta-critical trees and Yule or coalescent trees (already balanced). Values are averaged over 100 replicates. Shading depicts the 25\% to 75\% interquartile range. }
\end{figure}

Figure~\ref{fig:treeHeights} depicts the tree height of the original tree divided by the height of the decomposition tree produced by Algorithm~\ref{algo:decomp} for different tree distributions, see below. Note that, to compare like with like, we calculate the height of the original tree as the number of nodes {\em and} edges on the longest path from a root to a leaf. This is the size of a decomposition tree corresponding to the original pruning algorithm.  The improvement in likelihood computation times is, roughly, proportional to this ratio. 

There is a huge difference depending on the distribution we use to generate random trees. At one extreme, if we generate random caterpillar (completely unbalanced) trees then the original trees have height $n-1$ while the decomposition trees have height $O(\log n)$. With 7000 taxa that is a 1000-fold difference. 
At the other extreme, trees drawn from the Yule distribution (or equivalently, a coalescent or {\em equal-rate Markov} (ERM) distribution, see \cite{BlumFrancois06}) are already likely to be very balanced and we see at most a 2-3 fold difference with 7000 taxa. Trees generated using a beta-critical distribution, as introduced by \cite{aldous1996probability}, have  $O((\log n)^2)$ expected heights, giving a 7-fold expected height reduction with 7000 taxa. 
Between beta-critical  and caterpillar, there are trees drawn from a uniform  distribution (otherwise known as  {\em proportional to distinguishable arrangement} and PDA) have expected heights of around $O(\sqrt{n})$. This is still much larger than $O(\log n)$, and with 7000 taxa there is approximately a 30-fold difference on average.

\begin{figure*}
\centerline{\includegraphics[width=\textwidth]{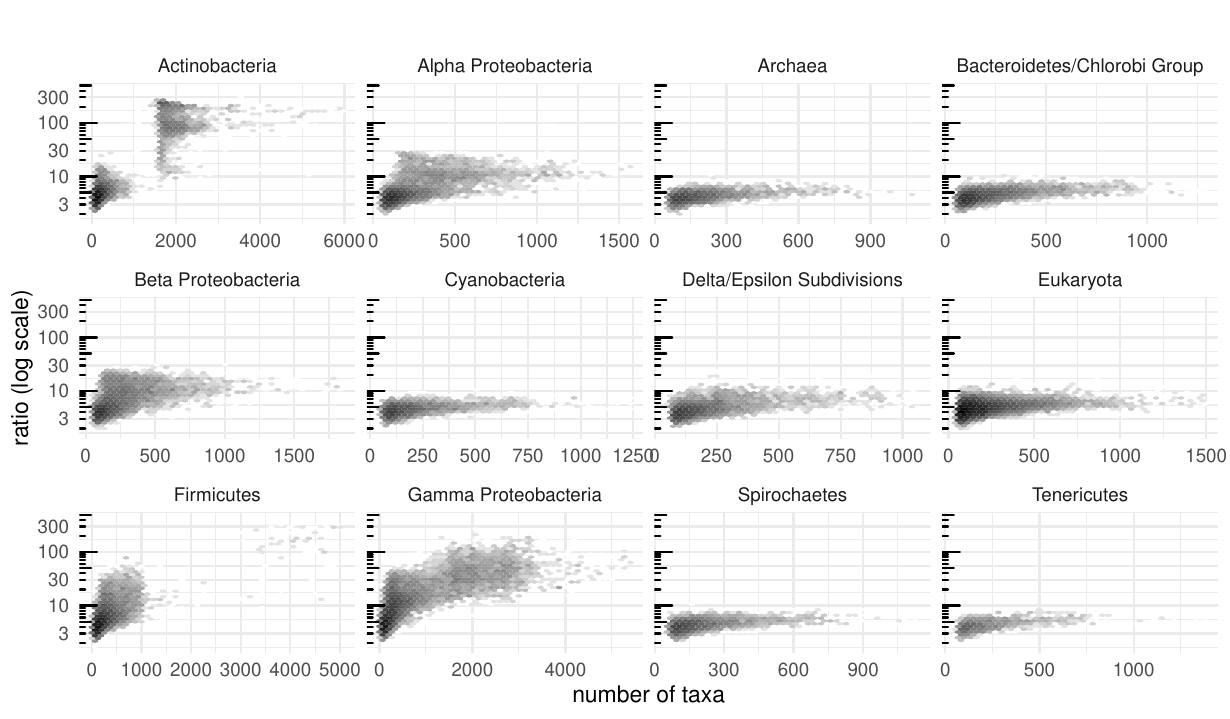}}
\caption{
A (log-scale) plot of the height of the original tree divided by the height of the decomposition tree produced by Algorithm 1 for different phyla of the Hogenome database \cite{penel2009databases}. \label{fig:heightHogenom}}
\end{figure*}

Which tree height distribution best matches `typical' phylogenies? This is a vexed, and ill-defined, question, and the answer depends very much on context. Theory tells us that gene trees within a single population are well described by a coalescent (Yule) distribution, so in that case we would not expect much difference in height between the decomposition tree and the original phylogeny. Others have observed that species trees are more unbalanced than Yule trees, but perhaps less unbalanced than uniform trees \citep{BlumFrancois06,Jones11}. The situation varies by taxa. Pathogen phylogenies, such as those for influenza, are often highly imbalanced, a phenomenon which can be explained by cross-immunity \citep{AndreasenSasaki06}. In Figure \ref{fig:heightHogenom}, we reported the height gains for different phyla of the Tree of Life \cite{penel2009databases}. This figure suggests that there is substantial potential for running-time improvements across different phyla, such as Alpha or Gamma Proteobacteria. 

\subsection{Computing likelihoods using the decomposition tree}

Like Felsenstein's pruning algorithm, we employ dynamic programming. We start with the base cases (components containing a single edge) and progressively update partial likelihoods for segments or clades formed by merging two others. The sequence of mergers is determined by the decomposition tree.

Consider first a component  $S$ consisting of a single edge $(u,v)$. There are two possible cases:
\begin{enumerate}
\item If $v$ is a tip then $S$ is a clade containing a single pendant edge. Let $s$ denote the observed state at $v$. Then, for all states $r$,
\begin{equation}   L_S(r) = \Pr[\x_v = s | \x_u = r].   \end{equation}   
If there is a missing state at tip $v$ then $L_S(r)=1$ for all $r$. 
\item If $u$ and $v$ are both internal then $S$ is a segment
containing a single internal edge and the partial likelihood is simply the transition probability
\begin{equation}   L_S(r,s) = \Pr[\x_v = s | \x_u = r].   \end{equation}   
\end{enumerate}

We now work up through the decomposition tree. Each time we visit a node we will have already visited its children. The partial likelihood functions are then updated according to the type of merger at each node (Fig.~\ref{fig:allMergers}).

\begin{enumerate}
\item[3.] If $S$ is formed from $S_1$ and $S_2$ with a type I merger, then for all $r$,
\begin{equation} L_S(r) = L_{S_1}(r) L_{S_2}(r).\label{eq:type1} \end{equation} 
\item[4.]  If $S$ is formed from $S_1$ and $S_2$ with a type II merger, then for all $r$ and $s$,
\begin{equation} L_S(r,s)=L_{S_1}(r)L_{S_2}(r,s).\label{eq:type2} \end{equation} 
\item[5.]  If $S$ is formed from $S_1$ and $S_2$ with a type III merger, then for all $r$ and $s$,
\begin{equation} L_S(r,s)= L_{S_1}(r,s) L_{S_2}(s).\label{eq:type3} \end{equation} 
\item[6.]  If $S$ is formed from $S_1$ and $S_2$ with a type IV merger, then for all $r$,
\begin{equation} L_S(r) = \sum_s L_{S_1}(r,s) L_{S_2}(s).\label{eq:type4} \end{equation} 
\item[7.]  If $S$ is formed from $S_1$ and $S_2$ with a type V merger, then for all $r$ and $s$,
\begin{equation} L_S(r,s)= \sum_t  L_{S_1}(r,t) L_{S_2}(t,s).\label{eq:type5} \end{equation} 
\end{enumerate}

\begin{figure}[ht]
\centerline{\includegraphics[width=0.7\textwidth]{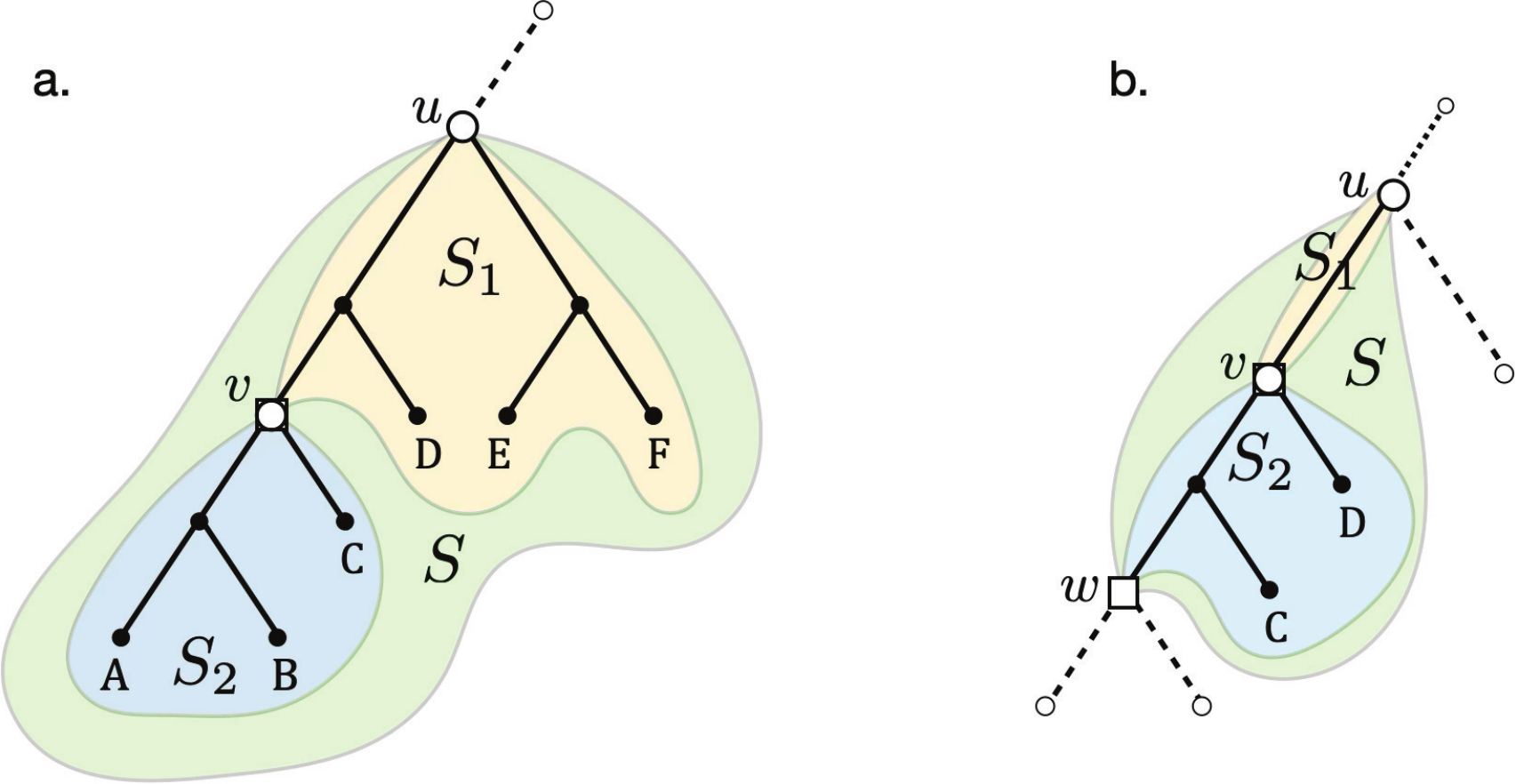}}
\caption{\footnotesize Examples of mergers: (a.) the clade $S_1$ and segment $S_2$ merge to form a clade $S$. (b.) the segments $S_1$ and $S_2$ merge to form a segment $S$.
}
\label{fig:mergeExample}
\end{figure}

Figure~\ref{fig:mergeExample} illustrates two of the ways that segments or clades can be merged to form other segments. 
In Figure~\ref{fig:mergeExample}a the bud of a segment $S_1$ coincides with the root of a clade $S_2$ and their union is a clade $S$. This is a type IV merger. Because of the Markov property of the model, we can factor the partial likelihood of the combined clade by summing over the state at the bud, giving
\begin{align}
L_{S}(r) & = \Pr[\X_{S_1},\X_{S_2} = \mbox{observed}|\x_u = r] \nonumber \\
& = \sum_s \Pr[\X_{S_1} = \mbox{observed},\x_v = s|\x_u = r] \nonumber \\
& \quad \quad \quad \times \Pr[\X_{S_2} = \mbox{observed}|\x_v = s]  \nonumber \\
& = \sum_s L_{S_1}(r,s) L_{S_2}(s).
\end{align}

In Figure~\ref{fig:mergeExample}b the bud of a segment $S_1$ coincides with the root of a segment $S_2$ and their union forms a segment $S$. This is a type V merger. By the same kind of argument,
\begin{align}
L_{S}(r,s) & = \Pr[\X_{S_1},\X_{S_2} = \mbox{observed}, \x_w = s |\x_u = r]  \nonumber \\
& = \sum_t \Pr[\X_{S_1} = \mbox{observed},\x_v = t|\x_u = r] \nonumber \\
& \quad \quad \quad \times \Pr[\X_{S_2} = \mbox{observed},\x_w = s |\x_v = t]  \nonumber \\
& = \sum_t L_{S_1}(r,t) L_{S_2}(t,s).
\end{align}
This calculation is just a multiplication of two matrices. As such, partial likelihood calculations for type V mergers involve approximately $r$ times as many arithmetic operations as other mergers, where $r$ is equal to the number of possible states in the alignment (so $r=4$ for nucleotides). When the number of type V mergers is large, this has an impact on the overall running time. As such, this method can be up to four times  {\em slower} than Felsenstein's pruning algorithm when computing the likelihood for a single site with a single processor. The running time improvements over the pruning algorithm kick in when we consider multiple sites, or multiple processors, and take advantage of the fact that the decomposition tree is balanced even if the original phylogeny is not, which will be the focus of the two next sections.

\subsection{Faster likelihood updates with LvD\label{sec:LvDUpdate}}

\begin{figure}[ht]
\centerline{\includegraphics[width=0.8\textwidth]{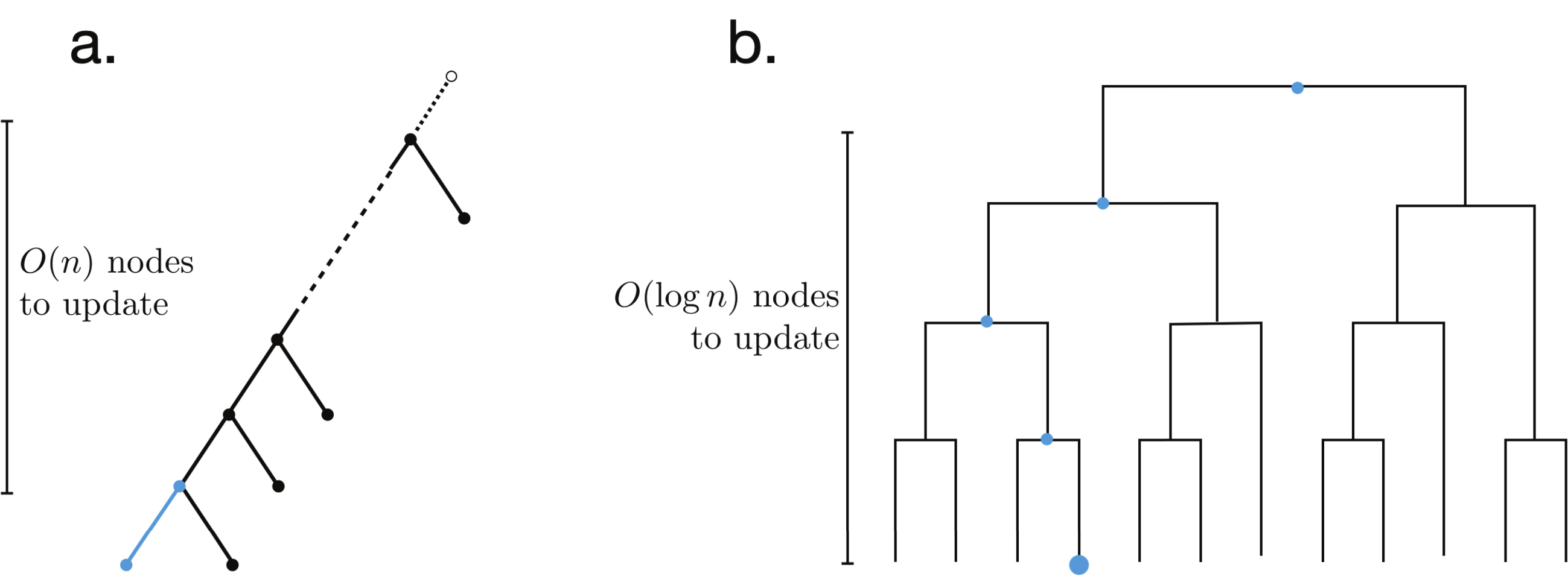}}
\caption{\footnotesize Updating partial likelihoods after a change of state at a tip. Suppose we change the state at the end of the highlighted pendant edge in (a). All partial likelihoods up to the root need to be updated, and there can be $n-1$ of these in the worst case (a caterpillar tree). In the decomposition tree (b) for that phylogeny, the same tree would require updates of only $O(\log n)$ nodes. The same applied for branch length changes.}
\label{fig:updatePath}
\end{figure}

Suppose that we have already computed partial likelihoods for a tree and we now wish to change the state for a single taxon, say from A to G. To update partial likelihoods we need to visit every node on the path from that taxon to the root, updating partial likelihoods along the way (Fig.~\ref{fig:updatePath}). In some cases only a few nodes need to be recomputed, however in many trees, the number of nodes will be linear in the number of leaves. 

In contrast, if we change the state for a single taxon in the decomposition tree then we only need to visit nodes in the decomposition tree on the path to the root, and since the decomposition tree has height $O(\log n)$ this means only $O(\log n)$ updates per change. This fact suggests an alternative strategy for computing the likelihood of a tree, given an alignment $A$, using a decomposition tree.
\begin{enumerate}
\item Calculate the partial likelihoods for the first site $A_1$.
\item For each subsequent site $A_{j+1}$, only update the partial likelihoods on the path from the root for taxa where the state for $A_{j+1}$ differs from that of $A_{j}$ .
\end{enumerate}
It makes sense to do some preprocessing on the site patterns before using this strategy, as suggested by \cite{Kosakovsky-PondMuse04}. We extract the distinct site patterns and sort them so that the average number of changes from one site to the next is (approximately) minimized. We implement a heuristic for the travelling salesman problem with sites as cities and distances given by the number of differences between different sites. Unlike \cite{Kosakovsky-PondMuse04} we do not repeat this calculation for every tree.

\begin{figure}[t]
    \centering
    \begin{subfigure}{0.45\textwidth}
        \centering
        \includegraphics[width=\textwidth]{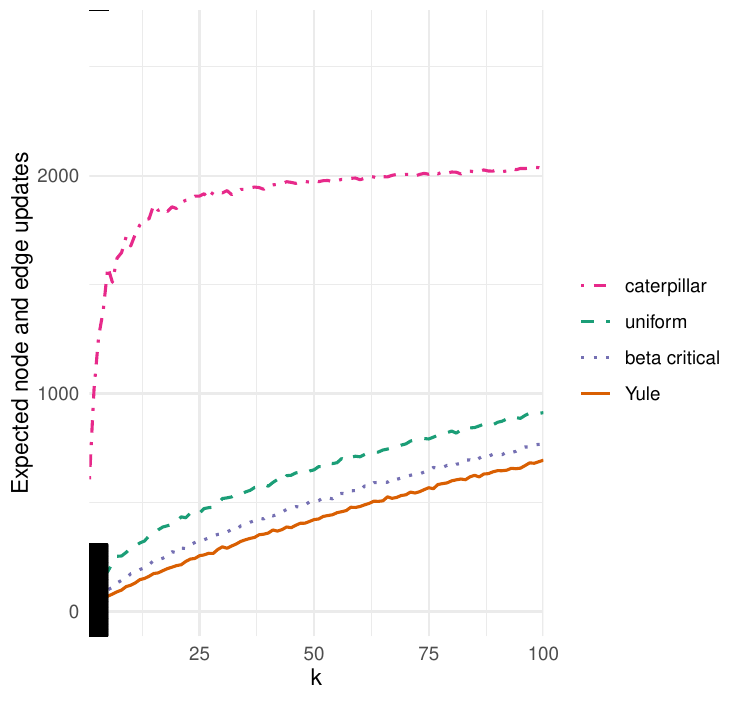}
        \caption{}
    \end{subfigure}
    \hfill
    \begin{subfigure}{0.45\textwidth}
        \centering
        \includegraphics[width=\textwidth]{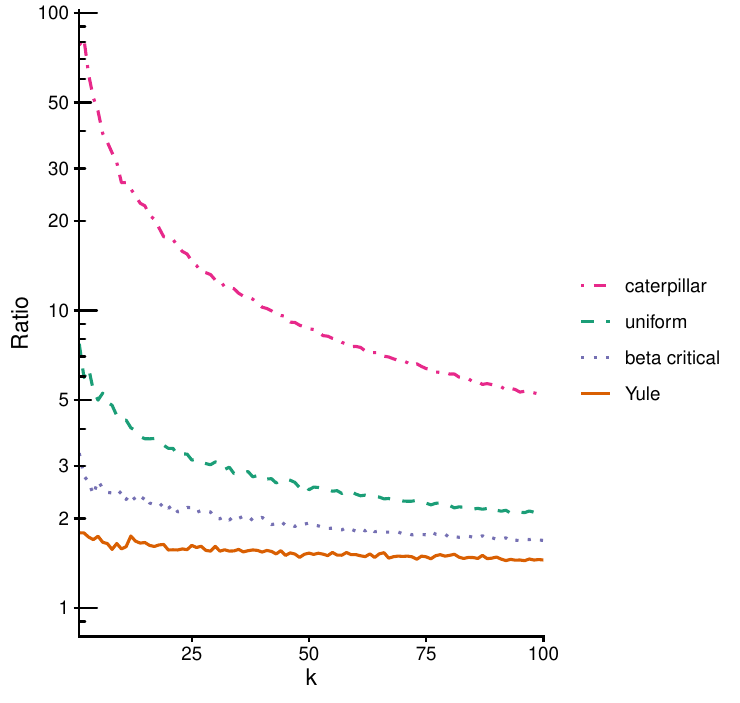}
        \caption{}
    \end{subfigure}
    \caption{ The expected number of node and edge updates required after changing $k$ states at the tips. (a) The number of changes required by the pruning algorithm, where we count updates at the top and bottom of each edge ancestral to a tip with changed state. (b) The ratio of the counts of updates required for the pruning algorithm and for the LvD algorithm. Figures were obtained by simulation $1000$ taxon trees at random from four distributions, and repeatedly simulated subsets of size $k$ of the set of taxa.     }
\label{fig:updatingCount}
\end{figure}

The speed improvements from this relatively simple strategy can be significant, but they depend  on:
\begin{enumerate}
\item The shape of the tree and the consequent reduction in tree height through LvD
\item The average number of tips which changed between consecutive sites in the minimal tour
\item The number of sites
\end{enumerate}
The more unbalanced the original tree is, the greater the speed-up with LvD. The more sites there are, and the less diverged the sequences are, the greater the speed-up, see Figure~\ref{fig:updatingCount}. In Figure~\ref{fig:updatingCount}(a) gives (simulated) numbers of partial likelihood updates required after changing $k$ random tip states. We examined trees with $1000$ taxa drawn from the distributions examined earlier. Figure~\ref{fig:updatingCount}(b) gives the ratio of updates when we compare the original tree to that used for LvD. In some cases (highly unbalanced trees, small number of changes each site) there are significant reductions. In other cases, such as trees already well-balanced, the improvement is more modest, though this will improve as the number of sites increases and the average number of differences between sites in a minimal tour decreases.

\subsection{A (theoretical)  $O(\log n)$ time parallel algorithm for likelihood}

With any algorithm, if there is a sequence of $k$ calculations which have to be carried out sequentially, one after the other, then no parallel algorithm 
will be able to improve on the time it takes to perform those $k$ calculations. With the standard pruning algorithm, the partial likelihood for a node cannot be determined until the partial likelihoods for its children have been computed. Hence the running time for any parallel implementation of the pruning algorithm is bounded below by the largest number of nodes on a path from a tip to the root.  In the worst case, this is $n-1$ in an $n$ taxa phylogeny. Rerooting can help, though the number of steps can still be linear in $n$.  Any parallel implementation of Felsenstein's pruning algorithm takes (worst case) linear $O(n)$ time to evaluate the probability of a single site, for any number of processors.

Consider the time taken to evaluate the probability of a single site when working with a balanced decomposition tree. The maximum length 
of a path from a node to the root is at most $O(\log n)$, irrespective of whether the original phylogeny was balanced or not.  With an unlimited number of processors we can first determine the partial likelihoods for all leaves, then all nodes with height one, then all nodes with height two, and so on. The total time taken will be $O(\log n)$.

The terminology of parallel computing distinguishes between two concepts, work and depth \citep{Blelloch96}. The {\em work} is the total amount of computation time taken, summed over all processors. In this case the work is $O(n)$ per site, for both the pruning algorithm and LvD, although this hides the fact that LvD will run slower if there are lots of Type V mergers. The {\em depth} or {\em span} is the fastest possible time that the computation can be carried out if the number of processors is unlimited. In this case, the depth of the pruning algorithm is (worst case) $O(n)$ while LvD is (worst case) $O(\log n)$. 

Brent's theorem \citep{Brent74}, a classical result in parallel computation, states that an algorithm with work $W$ and depth $D$ can be implemented on $p$ processors in time $O(W/p + D)$. As a consequence, if the number of processors exceeds $\frac{n}{\log n}$ then LvD takes $O(\log n)$ time. 
Unfortunately, the translation of this theoretical result to a practical speed up is far from automatic, and there is a lot of dependence on the particular computer architecture and type of parallelism. 

There is also a widely used strategy for implementing likelihood calculations in parallel: simply allocate different sites to different processors. For example, if there are 10 processors and 1000 sites (or site patterns) we might simply allocate 100 sites to each processor and then sum log likelihoods once they have completed. This approach can be combined with the LvD based updating strategies as described above. 

For now, we present the theoretical results here but focus on the non-parallel algorithms in our experiments. Work on parallelization is ongoing and will be the focus of a forthcoming paper where we will show the potential of LvD + parallelization, particularly for extremely large trees.

\section{LvD updating in practice - experimental results } \label{sec:experimental}

\subsection{Simulated data}

To assess the runtime improvements resulting from the LvD approach we implemented Felsenstein's pruning algorithm and the LvD updating approach within a common code base and compared running times on a large assortment of simulated and empirical data. 

For the experiments of simulated data, we proceeded as follows. We varied the number of taxa among 100, 1,000, and 10,000; the number of sites among 1,000, 10,000, and 100,000; and the tree height among 0.1, 0.01, 0.001, and 0.0001 expected substitutions per site. For each parameter combination, we generated 100 replicates. Trees were simulated under the same four distributions studied in Section~\ref{sec:LvDUpdate}, and sequence alignments were generated along these trees using the Jukes-Cantor (JC) model. We note that the choice of substitution model does not impact the relative running time of the likelihood algorithm, so we use Jukes-Cantor for simplicity. 

For each simulated tree and alignment we evaluated likelihoods using the pruning algorithm and LvD decomposition. We implemented the same updating strategy in both cases, the difference being in the algorithm used to update the partial likelihoods. To reduce implementation bias, we used the same code in both cases, the only difference being in how the decomposition tree was constructed. As we observe above, Felsenstein's pruning algorithm can be viewed as a special case of LvD but with a specially constructed decomposition tree.

\begin{figure}[t]
    \centering
    \begin{subfigure}{0.48\textwidth}
        \centering
        \includegraphics[width=\textwidth]{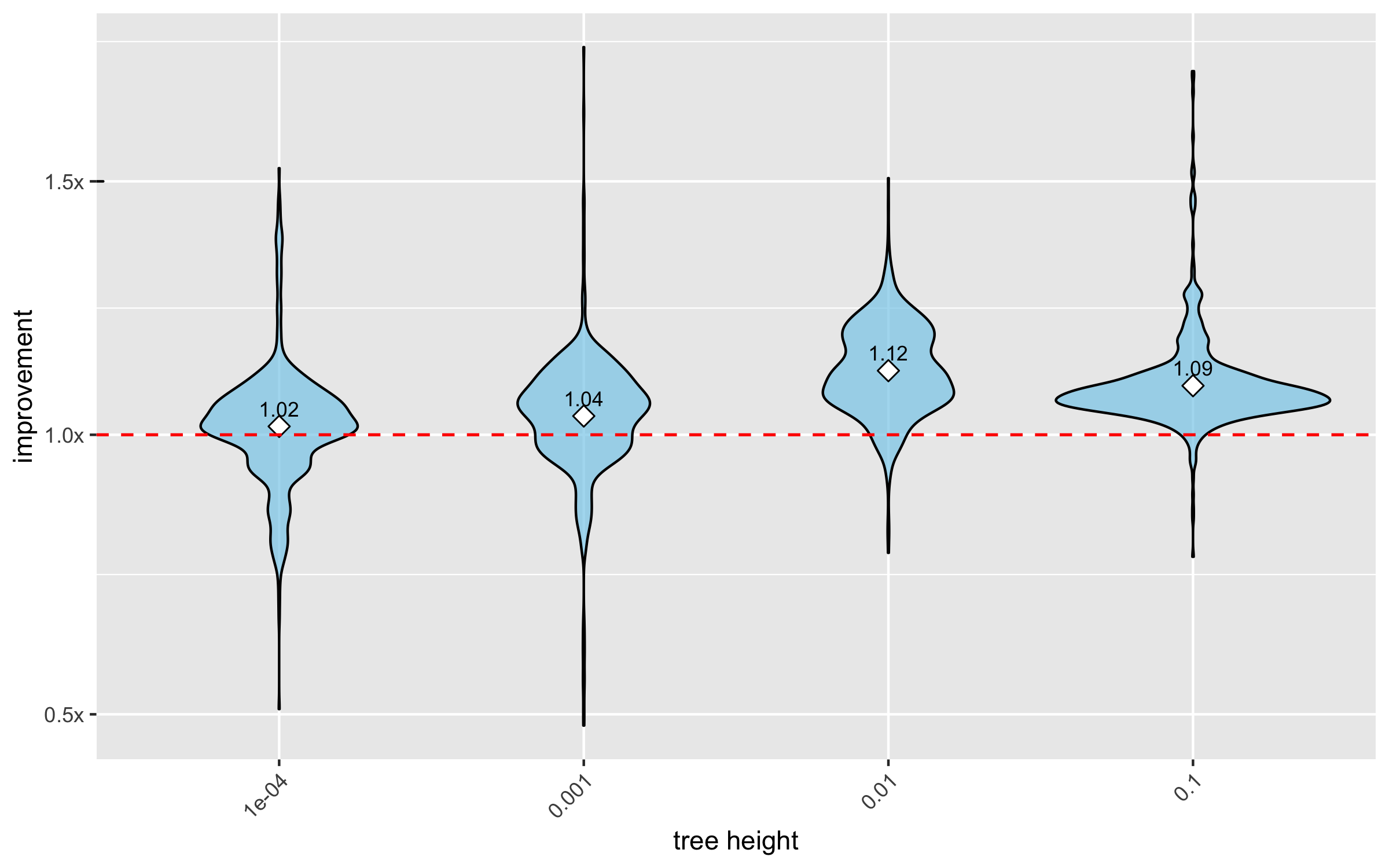}
        \caption{Yule}
        \label{fig:a}
    \end{subfigure}
    \hfill
    \begin{subfigure}{0.48\textwidth}
        \centering
        \includegraphics[width=\textwidth]{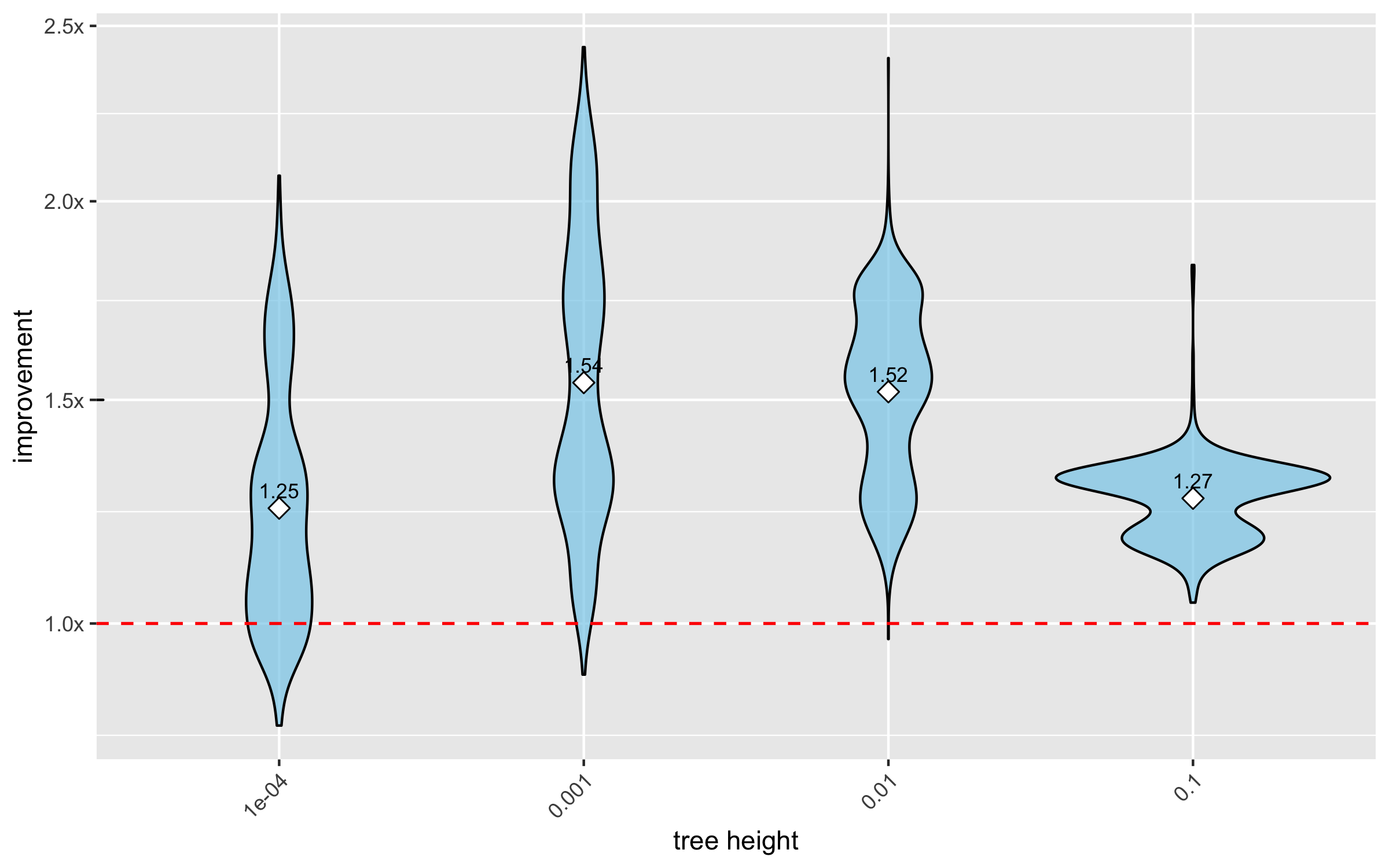}
        \caption{Beta critical}
        \label{fig:b}
    \end{subfigure}\\
        \begin{subfigure}{0.48\textwidth}
        \centering
        \includegraphics[width=\textwidth]{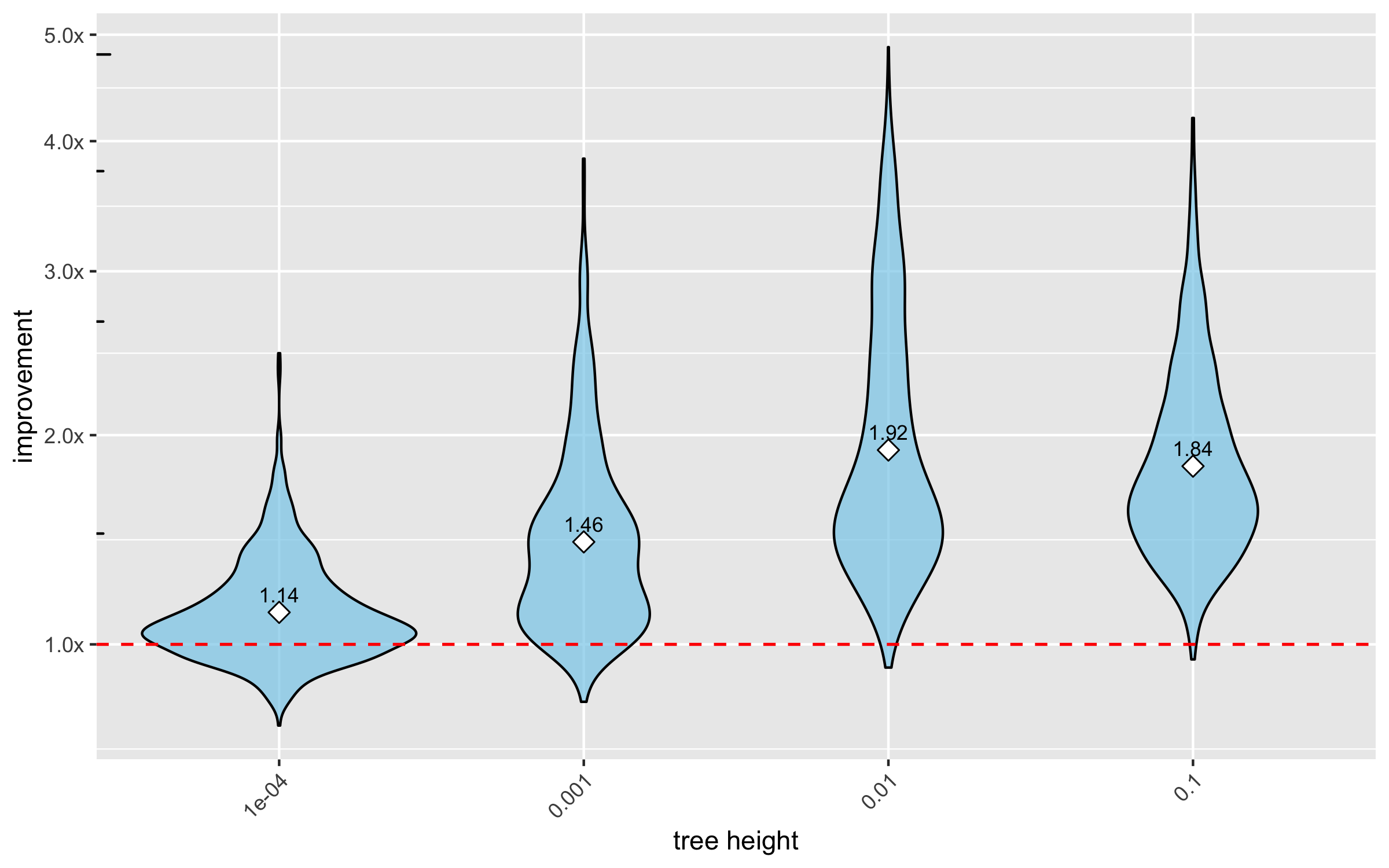}
        \caption{Uniform}
        \label{fig:c}
    \end{subfigure}
    \hfill
    \begin{subfigure}{0.48\textwidth}
        \centering
        \includegraphics[width=\textwidth]{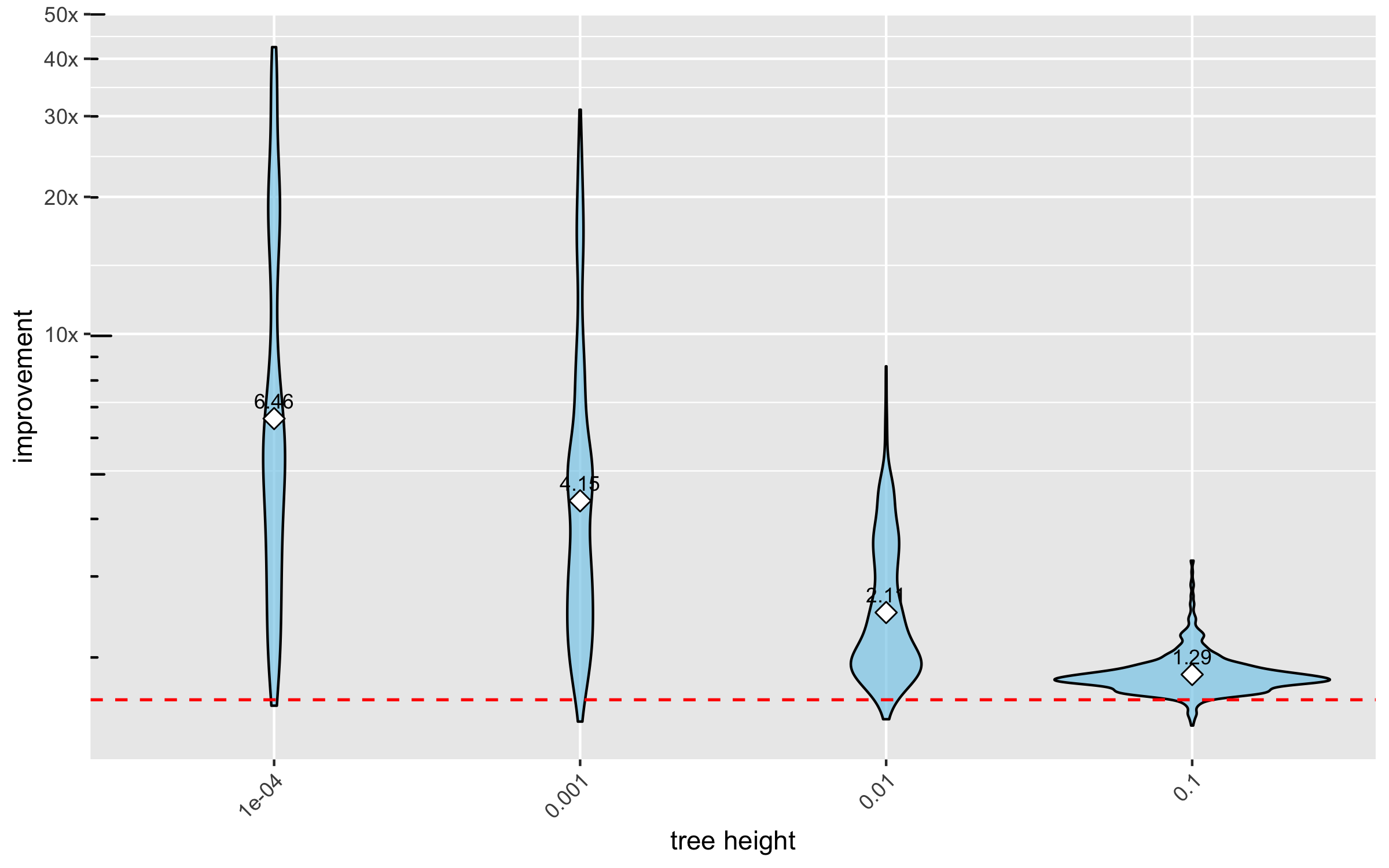}
        \caption{Caterpillar}
        \label{fig:d}
    \end{subfigure}\\
    \caption{Violin plots of the running-time improvements for  four tree distributions, as a function of tree height. Vertical axis indicate how much longer the pruning algorithm took compared to the LvD approach, displayed on a log scale.}
\label{fig:simRT}
\end{figure}

We report how many times faster the LvD approach is compared to the pruning algorithm. A value of $1$ (or less) indicates that there was no speed-up. Figure~\ref{fig:simRT} shows violin plots of the running-time improvements for the four tree distributions, as a function of tree height. The red dashed line indicates the level at which there is no improvement. Values are plotted on a log scale.

Overall the simulations indicate a modest to appreciable increase in speed with the new algorithm, with the increase being more significant the more the trees are unbalanced. When phylogenies are simulated from the Yule distribution then we see improvements ranging from between 2\% and 10\% on average. This makes sense as these phylogenies are already close to optimally balanced and little improvement is obtained through LvD. When phylogenies are simulated from the beta-critical distribution we see a 25\%-50\% improvement, depending on the tree height, while simulating from the uniform gives a 14\%-92\% average improvement, with speed-ups of 4-5 times not uncommon. 

Random caterpillar trees are the most unbalanced that we can get. For phylogenies drawn from this distribution we see speed-ups ranging from 650\% to 30\%, decreasing as the tree height increases. This dependency on tree height contrasts with the other distributions, however it has a simple explanation. The caterpillar trees we simulate are still clock-like, meaning that as the tree height increases we get extremely long branches and essentially randomized site patterns. If site patterns vary too much then we need to recalculate a linear number of partial likelihoods every site, undermining any improvement which can be obtained from clever updating. 

\begin{figure}[t]
    \centering
    \begin{subfigure}{0.48\textwidth}
        \centering
        \includegraphics[width=\textwidth]{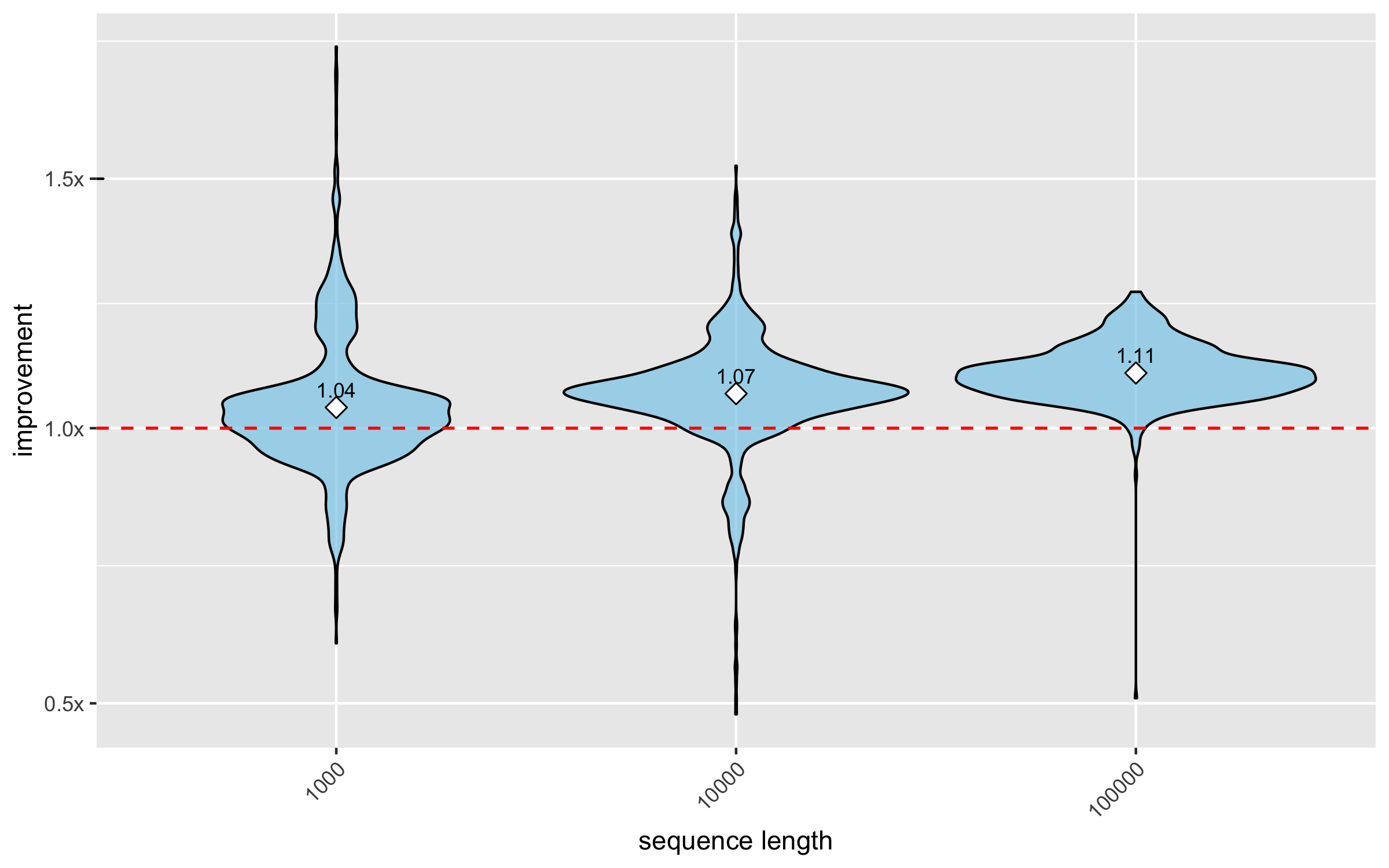}
        \caption{Yule}
        \label{fig:a2}
    \end{subfigure}
    \hfill
    \begin{subfigure}{0.48\textwidth}
        \centering
        \includegraphics[width=\textwidth]{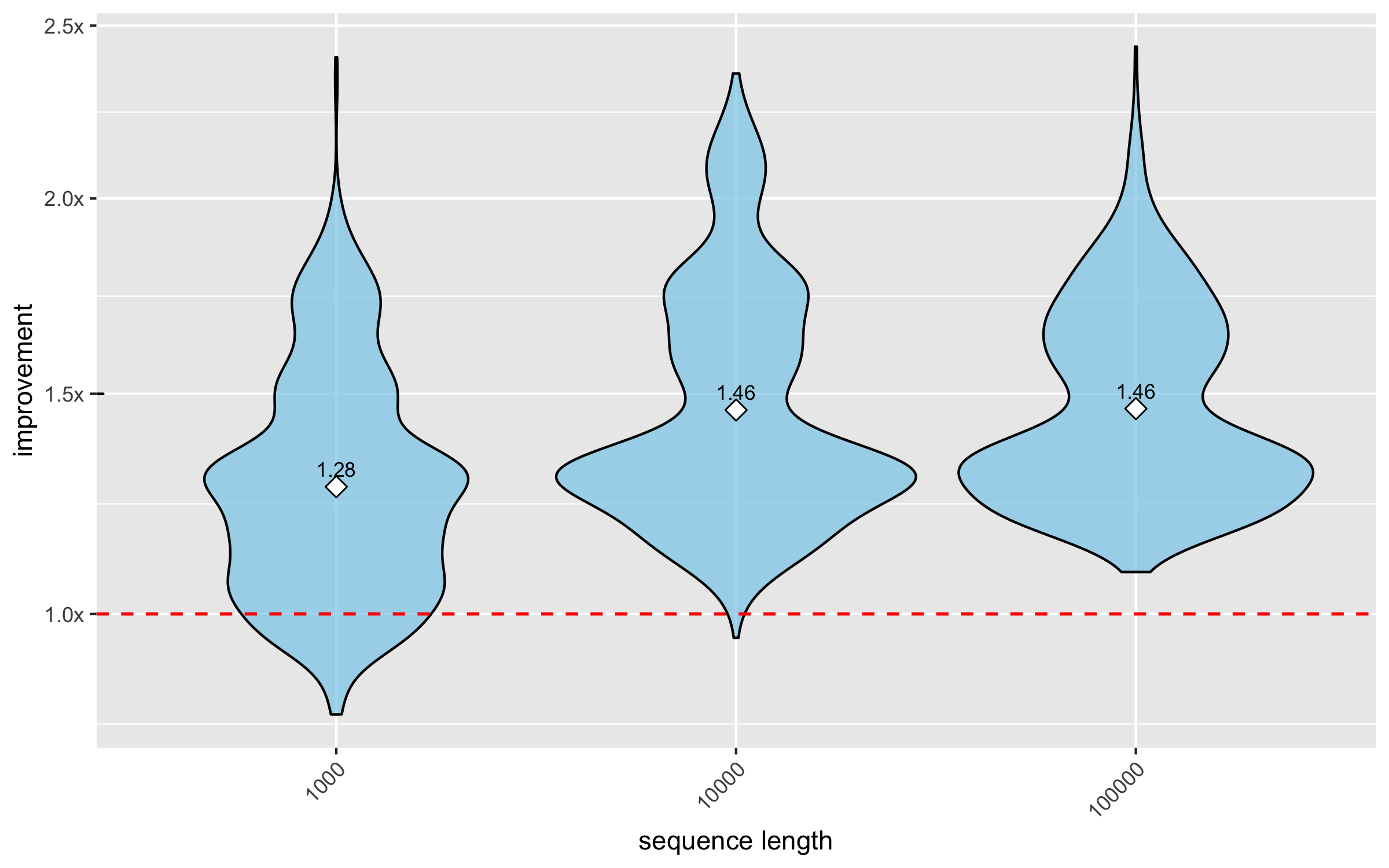}
        \caption{Beta critical}
        \label{fig:b2}
    \end{subfigure}\\
        \begin{subfigure}{0.48\textwidth}
        \centering
        \includegraphics[width=\textwidth]{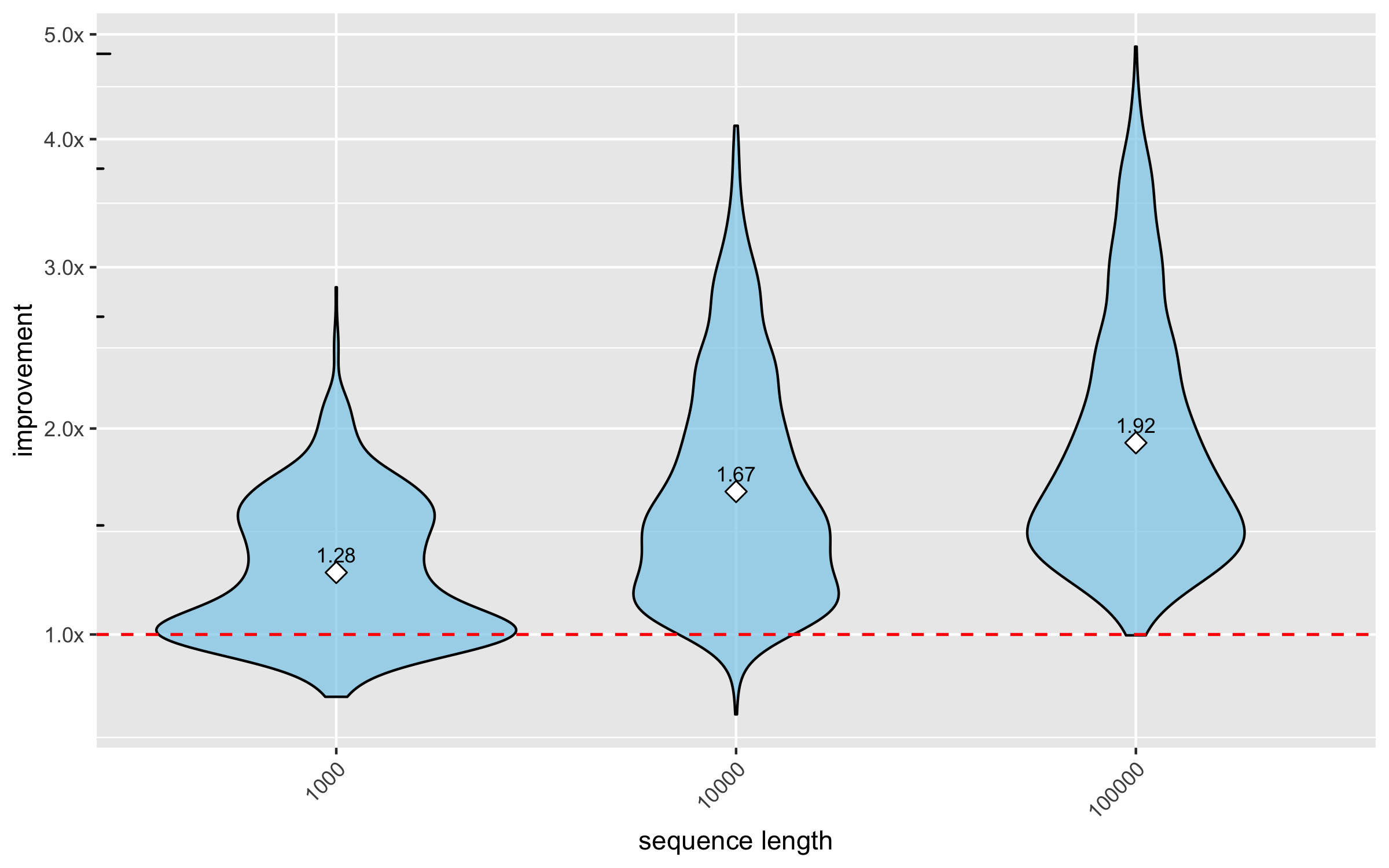}
        \caption{Uniform}
        \label{fig:c2}
    \end{subfigure}
    \hfill
    \begin{subfigure}{0.48\textwidth}
        \centering
        \includegraphics[width=\textwidth]{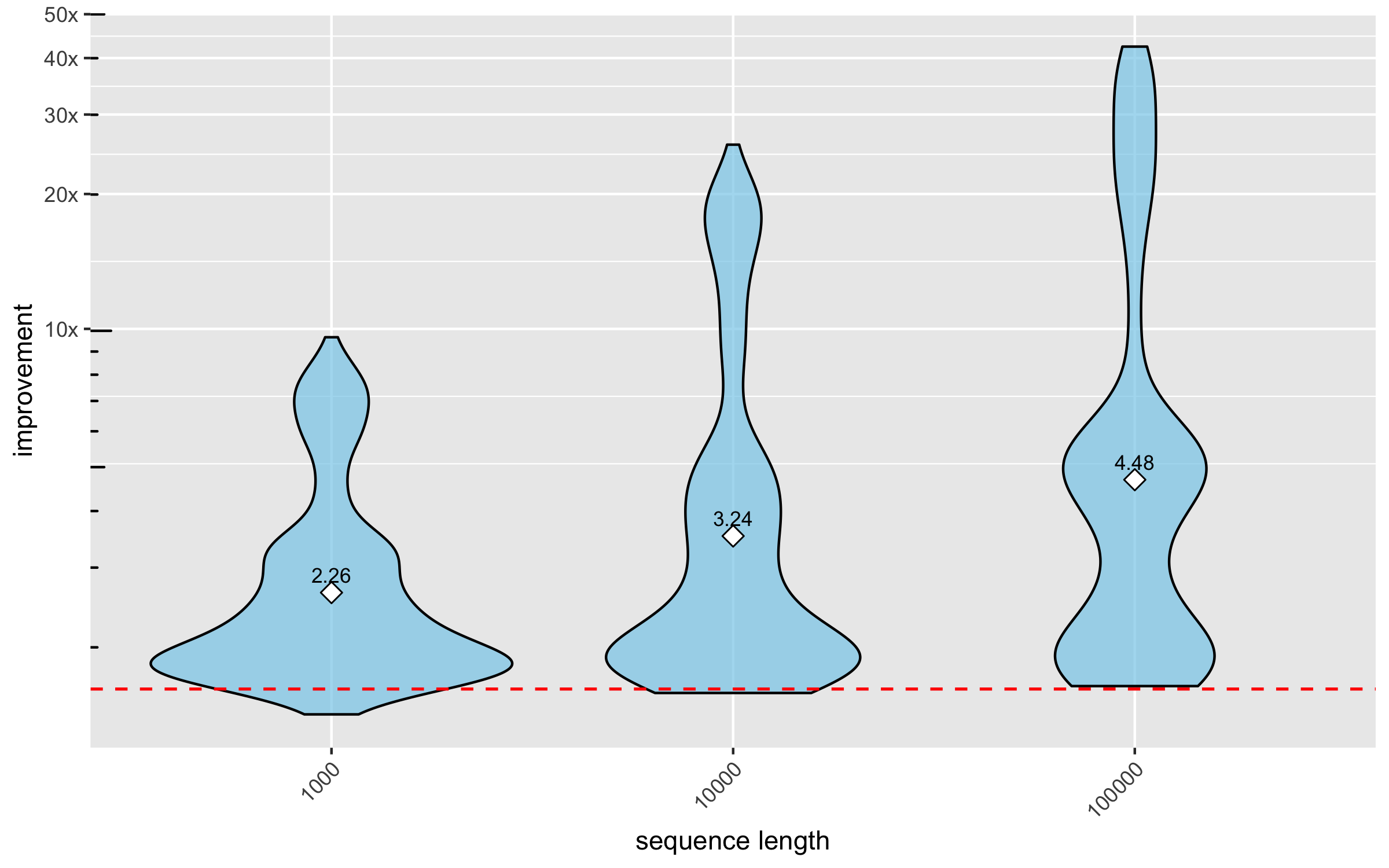}
        \caption{Caterpillar}
        \label{fig:d2}
    \end{subfigure}\\
    \caption{Violin plots of the running-time improvements for  four tree distributions, as a function of sequence length. Vertical axis indicate how much longer the pruning algorithm took compared to the LvD approach, displayed on a log scale.}
\label{fig:simRT2}
\end{figure}

Figure~\ref{fig:simRT2} is a plot of the same data but separated according to sequence length. Overall, we should expect faster relative running times per site as the sequence length increases due to repetition of high-frequency sites. This applies, however, to both the pruning algorithm and the LvD version. 

\subsection{Empirical data}

To explore the  performance of the LvD updating approach in practice, we also evaluated running-time improvements on five empirical data sets previously used to assess the performance of likelihood-based methods:
2000 pairs tree-alignment with more than 100 taxa of the RAxML Grove database \citep{10.1093/bioinformatics/btab863}, TreeBASE \citep[downloaded via the TreeBASEMirror \url{https://github.com/angtft/TreeBASEMirror}] {piel2009treebase}, the Gamma Proteobacteria and Eukaryota datasets of HOGENOM 
\citep{penel2009databases}  and an empirical data set used to test the performances of RAxML-NG v1.2 \citep{TogkousidisStamatakisEtal25}. 

\begin{figure}[ht]
\centerline{\includegraphics[width=0.8\textwidth]{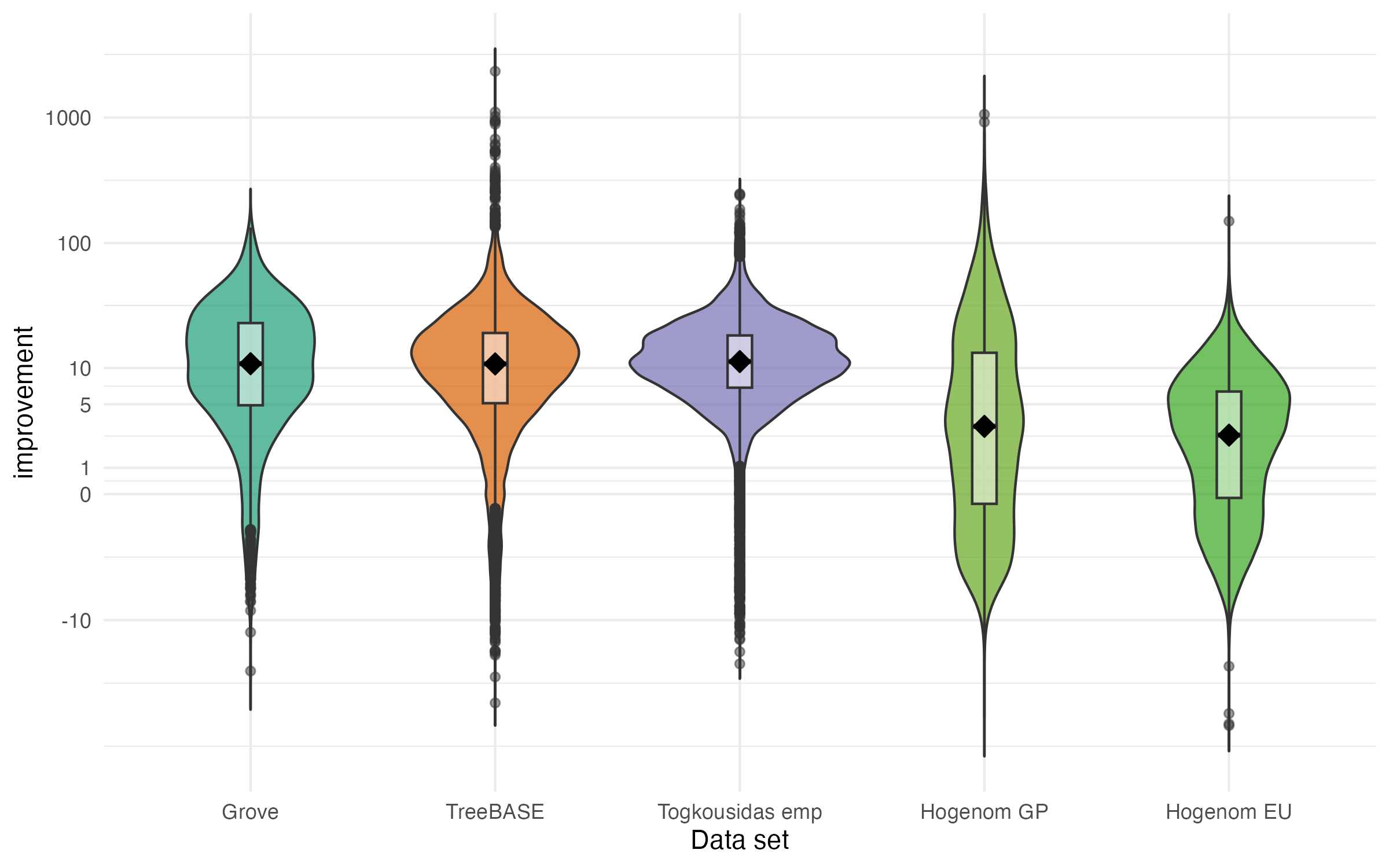}}
    \caption{Violin plots of the running-time improvements for  five tree datasets: RAxML Grove, TreeBASE, the empirical dataset of \cite{TogkousidisStamatakisEtal25}, and the Gamma Proteobacteria and Eukaryota subdatasets of HOGENOM.}
\label{fig:updating}
\end{figure}

The distribution of running time improvements for each data set are displayed in Figure~\ref{fig:updating}. Across this broad battery of tests, cases of negative improvement are negligible, while positive improvements typically correspond to speed-ups on the order of 20-30\%. Slightly lower speed-ups are observed for Eukaryota, which tend to exhibit more balanced tree shapes, more saturation and larger proportions of missing data.

Our experiments on these data sets, selected to capture the diversity of distributions observed in empirical data, demonstrate the strong potential of our new method. 
Although these results are encouraging, further evaluation is required to assess the final performance gains achievable in practice. In particular, our method must be integrated into state-of-the-art phylogenetic inference tools to fully quantify its impact when combined with the numerous algorithmic and implementation-level optimizations employed by software such as RAxML~\cite{Stamatakis2014RAxML}, PAUP~\cite{Swofford03} and IQ-TREE~\citep{Nguyen2015IQTREE}.

\section{Discussion} \label{sec:discussion}

We have presented a new algorithm for computing the likelihood of a phylogeny, arguably the first major modification of the established pruning algorithm in fifty years. The new approach uses a different scheme for decomposing the likelihood calculation, effectively replacing  dynamic programming on the phylogeny with dynamic programming on a new, balanced, tree structure.  Importantly, the new structure has {\em worst case} logarithmic height, even if the height of the original phylogeny is linear in the number of taxa. 

The new structures present multiple opportunities for more efficient computations, updates and parallelizations. In this paper, we have focused on two of these. The first is an efficient method to update likelihood calculations after changes in the branch lengths or leaf states. This in turn allows a reduction in the time taken to compute the likelihood of a tree, combining our updating algorithms with a site sorting approach developed by \cite{Kosakovsky-PondMuse04}. We note that the new algorithm requires no caching of partial likelihoods for multiple sites, meaning that the approach is extremely memory efficient. 

Experimentally, the computational gains from this application of the new structures give appreciable, if sometimes modest, improvements in overall running time. It is difficult to say what an `average' data set is, but overall we are seeing speed-ups in the order of 20-30\% on a wide range of alignments---in some cases lower, in other cases much higher. This might seem like an awful lot of effort for a 20-30\% speed-up, but there are two important factors to appreciate. Firstly, this is a speed-up of the {\em core} algorithm, the most computationally intensive component of any phylogenetic likelihood analysis, and the improvements we make will be {\em in addition} and {\em complementary} to the wide range of algorithmic tricks and techniques already implemented. Secondly, the method we describe is only one way that the new decomposition structures could be exploited, and a fairly simple one at that. There is considerable scope for optimizations of this approach, and indeed of more sophisticated strategies to take advantage of the new framework. 

In addition to efficient updating, we also observe that the new structures could, in principle, lead to new and more efficient parallelisation strategies for computing the likelihood. Our contribution here is more on the theoretical side, but it is nevertheless a contribution with practical implications. We show that the likelihood of a phylogeny can be computed with $O(\log n)$ depth, that is, $O(\log n)$ time on an unbounded number of processors. The importance of this fact rests on the fact that depth provides an absolute and immutable barrier to efficient parallel computation, irrespective of the amount of clever software engineering. We have improved the worst case depth from linear (in the number of taxa) to logarithmic. While the development of any practical parallel algorithm requires a great deal of bespoke engineering we have at least created new opportunities for improved efficiency, especially for phylogenies with huge numbers of taxa. 

There is, of course, still a great deal of work to do exploring the opportunities created by LvD. There are also some significant caveats. The first is that while phylogenies have {\em worst case} height which is linear in the number of taxa, the bulk of phylogenies in the various databases are already far more balanced. It is still not completely clear whether this is a product of some bias in phylogenetic inference, or peculiarities in taxon sampling. We are also not sure whether this applies both to the inferred phylogenies to a greater extent than to all the many suboptimal phylogenies encountered during a likelihood search or sampling algorithm. In any case, if a phylogeny is already quite balanced then the gains due to LvD will be modest, irrespective of the strategy applied. 

The second caveat is that the algorithm for computing the likelihood of a single site will, in general, run slower than the existing pruning algorithm, all other factors being equal. The reason is that updating partial likelihoods of a component after a type V merger requires the multiplication of two $r\times r$ matrices, whereas the pruning algorithm, and all other mergers, require at most the multiplication of a matrix by a vector. Because of this, the LvD algorithm can take four times longer for nucleotide alignments, in the worst case. The idea, of course, is that this cost is offset by the improved updating or parallelization gains, though we found that when phylogenies were already quite well-balanced the extra computation time did make an impact.  As well, the additional computation required for type V mergers will be even more significant if we move from nucleotide to protein or codon likelihoods. For these data, it will be necessary to modify the LvD algorithm in order to see any practical benefits. 

A number of open problems remain, or put otherwise, there are multiple opportunities for future improvements. Our algorithm for constructing the decomposition table optimizes for overall height, however it may be prudent to allow a small increase in height in return for a reduction in type V merges. As well, we have not considered updating the LvD structures after a change in the underlying phylogeny. The algorithm for constructing the decomposition table takes only linear time, so can be executed after each change in the phylogeny with little additional overhead. More important is whether it might be practical to update the partial likelihood values cached after such a change. Finally, while the approach we apply to updating has the advantage of low memory requirements, a careful use of caching might in fact lead to substantial real improvements in running time, carefully implemented.

\section{Acknowledgements}
We would like to thank Simon Penel for having provided the empirical tree distributions depicted in Figure~\ref{fig:heightHogenom}. 
This research was supported by University of Otago Science division research funding.

\bibliographystyle{abbrvnat}
\bibliography{lvd-manuscript}

\appendix 
\section{Decomposition tree proofs \label{sec:decompAlgo}}

Here we prove the correctness of Algorithm~\ref{algo:decomp}: that it returns a valid decomposition tree, that tree has $O(\log n)$ height, and that the algorithm can be run in linear $O(n)$ time. We do this by proving (or referencing) a sequence of smaller results.\\

(1) {\em At the beginning of  iteration $j$ of the main loop, $\{j,j+1,\ldots,k\} \subseteq M$}\\
When $j=1$, $k=m$ and $M = \{1,\ldots,m\}$ so the statement holds. In each iteration $j>1$ we possibly remove $i$ and $j$ from $M$ and add element $k$. Whether or not there is a merger in iteration $j$, the statement holds at the beginning of the subsequent iteration.\\

(2) {\em At the beginning of  iteration $j$ of the main loop,  $L[i] \leq L[j]$ for all $1 \leq i \leq j < k$.}\\
Initially, $L[i]=1$ for all $1 \leq i \leq m = k-1$. Suppose that we are at iteration $j$. For each $\ell$ such that $j<\ell<k$ we have that $S[\ell]$ was formed by merging components $S[i']$ and $S[j']$ for some $1 \leq i' < j' < j$. For each we have $L[\ell] = L[j'] + 1 \leq L[j]+1$. If we carry out a merger this iteration then we add a segment $S[k]$ with $L[k] = L[j] + 1 \geq L[\ell]$ for all $\ell = 1,\ldots,k-1$. Hence $L[1] \leq L[2] \leq \cdots \ldots L[k]$. \\

(3) {\em At the end of iteration $j$ of the main loop there no pairs $(i,i')$ such that $i,i' \in M$ such that $i < i' \leq j$ and $S[i] \cup S[i']$ forms a valid merger.}\\
Suppose that there was such a pair. Since we did not merge $S[i']$ at iteration $i'$ we must have merged $S[i]$ previous to iteration $i'$, a contradiction.\\

(4) {\em Let $T$ be a binary phylogeny with edge set $E(T)$.  Let $\sM$ be a collection of segments and clades which partition $E(T)$. Then there is a set of at least $\frac{|\sM|}{4}$ non-overlapping pairs such that $A,B \in \sM$ and $A \cup B$ is a segment or clade of $T$. }\\
This claim is a restatement of Proposition~3 in \cite{BryantScornavacca19}, albeit with slightly different terminology. The term `segment' in  \cite{BryantScornavacca19}  refers to a component with {\em at most} two boundaries (rather than {\em exactly} two boundaries).\\

(5) {\em At the conclusion of the algorithm, the decomposition tree has $O(\log n)$ height.}\\
Let $j_0 = 0$ and for $\ell = 1,2,\ldots$ let $j_\ell = \max\{j:L[j]=\ell\}$ and $n_\ell = j_\ell - j_{\ell-1}$. Hence $L[j]=\ell$ exactly when $j_{\ell-1} < j \leq j_\ell$. Let $M_\ell$ denote the value of $M$ at the conclusion of iteration $j_\ell$ of the main loop and let $m_\ell = |M_\ell|$. Hence $M_1 = \{1,\ldots,m\}$. The clades and segments in $\sM_\ell = \{S[i]:i \in M_\ell\}$ partition  $E(T)$, so by (4) there is a set $P_\ell$ of at least $\frac{|\sM_\ell|}{4} = \frac{|\sM_\ell|}{4}$ disjoint pairs in $\sM_\ell$  which can be merged. By (3), for each of these pairs $S[i],S[i'] \in P_\ell$ must involve a segment with index greater than $j$.

During the iterations $j=j_\ell+1,\cdots,j_{\ell+1}$ we merge all at least one component from each pair in $P_\ell$. As each merger involves two components, there are at least $\frac{|P_\ell|}{2}$ such mergers, one for each $j$ from $j_\ell+1$ to $j_{\ell+1}$. After these mergers there are $m_\ell - 2n_{\ell+1}$ elements left in $M$. We therefore have
\begin{align*}
n_1 & = m\\
m_1 & = m\\
n_{\ell+1} & \geq m_{\ell}/8 \\
m_{\ell+1} & = m_\ell - 2 n_{\ell + 1}.
\end{align*}
This gives
\[m_{\ell+1}  \leq \frac{3}{4} m_{\ell} \leq \left(\frac{3}{4}\right)^\ell m.\]
The total height $h$ therefore satisfies 
\[\left(\frac{3}{4}\right)^{h-1} m \geq 1\]
so that $h \leq \frac{\log(m)}{\log(4/3)}$ which is $O( \log n)$. We do not claim that this bound is tight.\\

(6) {\em The Algorithm can be implemented to run in linear time (in $n$).}\\
n order to achieve $O(n)$ runtime we never explicitly construct the sets of edges in each component. Instead, for each component encountered we store a pointer to the root, a pointer to the bud (or $\nul$ if the component is a clade), and flags indicating whether the component has been paired up in the current level. We also store a list of boundary nodes for the currently maximal components with links to the components they are boundaries for. These tables are updated in constant time after each merger. 

From Claim $1$, the number of boundary nodes examined during each iteration of the algorithm is linear in the number of components merged during that iteration. Hence the total running time is linear in the number of nodes of the decomposition tree, which is $O(n)$.\\

\end{document}